\documentclass[aps,nofootinbib,superscriptaddress
]{revtex4-2}
\usepackage{amsmath,amssymb,graphicx,epsfig,color,xcolor,ulem,cancel,datetime,dcolumn,bm}
\usepackage{graphicx}
\usepackage{amsmath}
\usepackage{amsfonts}
\usepackage{amssymb}

\begin{document}

\title{Catastrophic Emission of Charges from Near-Extremal Nariai Black Holes}

\author{Chiang-Mei Chen} \email{cmchen@phy.ncu.edu.tw}
\affiliation{Department of Physics, National Central University, Chungli 32001, Taiwan}
\affiliation{Center for High Energy and High Field Physics (CHiP), National Central University, Chungli 32001, Taiwan}

\author{Chun-Chih Huang} \email{makedate0809@gmail.com}
\affiliation{Department of Physics, National Central University, Chungli 32001, Taiwan}

\author{Sang Pyo Kim} \email{sangkim@kunsan.ac.kr}
\affiliation{Department of Physics, Kunsan National University, Kunsan 54150, Korea}
\affiliation{Asia Pacific Center for Theoretical Physics, Pohang 37673, Korea}
\affiliation{Helmholtz-Zentrum Dresden-Rossendorf, Bautzner Landstra\ss e 400, 01328 Dresden, Germany}

\author{Chun-Yu Wei} \email{weijuneyu@gmail.com}
\affiliation{Department of Physics, National Central University, Chungli 32001, Taiwan}

\date{\today}

\begin{abstract}
Using both the in-out formalism and the monodromy method, we study the emission of charges from near-extremal charged Nariai black holes with the black hole event and cosmological horizons close to each other, whose near-horizon geometry is $\mathrm{dS}_2 \times \mathrm{S}^2$. The emission becomes catastrophic for a charge with energy greater than its chemical potential, whose leading exponential factor increases inversely proportional to the separation of two horizons.
This effect may prevent near-extremal Nariai black holes with large charges that evaporate dominantly through the charge emission from evolving to black holes with a naked singularity, in analog to near-extremal RN-dS black holes that have the Breitenlohner-Friedman bound, below which they become stable against Hawking radiation and Schwinger effect of charge emission.
The near-extremal Nariai black holes with small charges, which are close to near-extremal Schwarzschild-dS black holes, emit dominantly charge-neutral particles and evolve to black holes with increasing charge to mass ratio. We illuminate the origin of the catastrophic emission in the phase-integral formulation and monodromy method by comparing near-extremal charged Nariai black holes with near-extremal RN-dS black holes.
\end{abstract}

%% REVTEX4
%\pacs{}

\maketitle
%\tableofcontents

%%%%%%%%%%%%%%%%%%%%%%%%%%%%%%%%%%%%%%%%%%%%%%%%%%%%%%%%%%%%%%%%%%%%%%
\section{Introduction}
%%%%%%%%%%%%%%%%%%%%%%%%%%%%%%%%%%%%%%%%%%%%%%%%%%%%%%%%%%%%%%%%%%%%%%
The black hole horizon and the cosmological horizon endow the spacetime geometry with interesting quantum field properties since both horizons emit Hawking radiation and Gibbons-Hawking radiation~\cite{Gibbons:1977mu}. The evaporation of all species of particles from a Schwarzschild or a Kerr black hole in the de Sitter (dS) space has been studied~\cite{Gregory:2021ozs}. Charged black holes in the de Sitter (dS) space, Reissner-Nordtr\"{o}m-dS (RN-dS) black holes, exhibit much rich structure due to the existence of both horizons emitting radiations~\cite{Romans:1991nq}. For a fixed dS radius, depending on the charge to mass ratio, the RN-dS black holes can have at most three horizons: the inner (Cauchy) horizon, the outer (black hole event) horizon, and the cosmological horizon. Then, they have two extremal limits: the near-extremal RN black holes and the Nariai black holes. The near-extremal RN black holes, where the inner horizon and the event horizon are close to each other, have a near-horizon geometry of $\mathrm{AdS}_2 \times \mathrm{S}^2$, in which a quantum field equation can be solved in terms of special function due to an enhanced symmetry~\cite{Bardeen:1999px}. Recently, two of the authors (CMC and SPK) have studied the emission of charges from near-extremal Reissner-Nordstr\"{o}m (RN) black holes and Kerr-Newmann (KN) black holes in the dS space~\cite{Chen:2020mqs, Chen:2021jwy}. The black hole thermodynamics has been studied in the near-extremal and extremal limit of RN-dS black holes~\cite{Castro:2022cuo}.

On the other hand, Nariai black hole is the coincidence limit of the event horizon and the cosmological horizon. Both horizons emit radiations with the Hawking temperature and the Gibbons-Hawking temperature, respectively. Except for the ``lukewarm" limit, the RN-dS black hole cannot remain a thermal equilibrium since the black hole temperature is in general higher than the Gibbons-Hawking temperature. However, as the two horizons get close to each other, the gap between the Hawking temperature and the Gibbons-Hawking temperature narrows but each temperature diminishes because of the near extremality. This means that Hawking radiation and Gibbons-Hawking radiation are exponentially suppressed. In contrast, the near-extremal limit of a charged black hole still has an electric field between the two horizons and is an analog of conductor of two spherical shells, in which Schwinger effect of pair production is the main mechanism for the emission of charges.

The near-extremal charged Nariai black hole has a near-horizon geometry of $\mathrm{dS}_2 \times \mathrm{S}^2$ in comparison to $\mathrm{AdS}_2 \times \mathrm{S}^2$ of the near-extremal RN-dS black hole. The Schwinger effect in $\mathrm{dS}$ space~\cite{Garriga:1994bm, Kim:2008xv} differs from that in $\mathrm{AdS}$ space~\cite{Pioline:2005pf, Kim:2008xv}. In this paper we will study the emission of charges from near-extremal Nariai black holes and show that the emission becomes catastrophic as the distance between two horizons draws closer and closer. This is interesting because Hawking radiation and Gibbons-Hawking radiation are exponentially suppressed due to their small temperatures, but the spontaneous pair production via the Schwinger mechanism becomes a dominant channel for charge emission.

Using the near-horizon geometry $\mathrm{dS}_2 \times \mathrm{S}^2$ of near-extremal Nariai black holes in the Einstein-Maxwell theory~\cite{Ortaggio:2002bp}, we solve the field equation for a charged scalar field, and properly selecting the in- and the out-vacua in the region between two horizons and the region exterior to the cosmological horizon, we find the mean number of pair production in the in-out formalism. Besides, we apply the monodromy method to the Riemann P-function~\cite{Chen:2022hpe} which includes the field equation in both the outer region and in-between region, and find the same mean number for pair production. The monodromy method uses local behaviors of the wave function at singular points and finds the connection matrix for scattering problem~\cite{Castro:2013kea, Castro:2013lba}.

We show that Nariai black hole cannot remain quantum mechanically stable since there is no Breitenlohner-Freedman (BF) bound that guarantees stability against the emission of charges for (near-) extremal RN-dS black holes. This implies that near-extremal Nariai black holes can evaporate either to RN-dS black holes or black holes with one horizon and a naked singularity, which depend on the ratio of charge to mass~\cite{Montero:2019ekk}.
The catastrophic emission of charges with energy greater than chemical potential evolves the near-extremal Nariai black holes to nonextremal RN-dS black holes. Remarkably the Schwinger emission of charges with large charge to mass ratio, which is a possible physical process, violates the cosmic censorship conjecture.
Furthermore, we confirm that the leading emission Boltzmann factor from Nariai black holes exhibits a universal thermal interpretation with an effective temperature which is determined by the Unruh temperature for charge acceleration and the Gibbons-Hawking temperature associated with the dS radius~\cite{Cai:2014qba}.

The organization of this paper is as follows. In Sec.~\ref{sec II} we study the geometry of Nariai black hole and extend it to a near-extremal Nariai black hole with the event and cosmological horizons close to each other. In Sec.~\ref{sec III} we find the emission formulae of charges from near-extremal Nariai black hole, both in the region between two horizons and the region outer to the cosmological horizon. In Sec.~\ref{sec IV} we compare the emission from the near-extremal Nariai and RN-dS black holes. In particular, the physical reason for catastrophic emission is explained. In Appendix~\ref{app_A} we explain the boundary condition for a quantum field in a timelike region and a spacelike region, which is used to define the in- and the out-vacua for the in-out formalism. In Appendix~\ref{app_B}, using the Riemnann P-function, we recapitulate the monodromy method to find the mean number for pair production, which includes the emission from near-extremal Nariai black holes.

%%%%%%%%%%%%%%%%%%%%%%%%%%%%%%%%%%%%%%%%%%%%%%%%%%%%%%%%%%%%%%%%%%%%%%
\section{Nariai black holes} \label{sec II}
%%%%%%%%%%%%%%%%%%%%%%%%%%%%%%%%%%%%%%%%%%%%%%%%%%%%%%%%%%%%%%%%%%%%%%
The Reissner-Nordstr{\"o}m-de Sitter (RN-dS) solution of a charged black hole is\footnote{The geometric units of $c = \hbar = 4 \pi \epsilon_0 = G = 1$ are used, where time, length, mass, and charge are all dimensionless. Time, length, mass, charge, and energy measured in the Planckian units recover the SI units.}
\begin{equation} \label{eq_RNdS}
ds^2 = - f(r) dt^2 + \frac{dr^2}{f(r)} + r^2 d\Omega_2^2, \qquad A_{[1]} = \frac{Q}{r} dt,
\end{equation}
where the lapse function can have maximally three positive horizons and a negative root:
\begin{equation}
f(r) = 1 - \frac{2 M}{r} + \frac{Q^2}{r^2} - \frac{r^2}{L^2} = - \frac{(r - r_-) (r - r_+) (r - r_c) (r + r_- + r_+ + r_c)}{L^2 r^2}.
\end{equation}
The three parameters (hairs) are the mass $M$, charge $Q$, and cosmological constant $\Lambda = 3/L^2$ ($L$ being the dS radius). The inner (Cauchy) horizon $r_-$, the outer (event) horizon $r_+$, and the cosmological horizon $r_c$ in increasing order, are related to the physical parameters as
\begin{eqnarray}
L^2 &=& r_+^2 + r_-^2 + r_c^2 + r_+ r_- + r_+ r_c + r_- r_c,
\\
M &=& \frac{(r_+ + r_-) (L^2 - r_+^2 - r_-^2)}{2 L^2} = \frac{(r_+ + r_c) (L^2 - r_+^2 - r_c^2)}{2 L^2},
\\
Q^2 &=& \frac{r_+ r_- (L^2 - r_+^2 - r_-^2 - r_+ r_-)}{L^2} = \frac{r_+ r_c (L^2 - r_+^2 - r_c^2 - r_+ r_c)}{L^2}.
\end{eqnarray}
The associated Hawking temperature, entropy, and electric potential at the black hole horizon are given by
\begin{eqnarray}
&& T_H = \frac{f'(r_+)}{4 \pi} = \frac{(r_+ - r_-) (L^2 - 3 r_+^2 - 2 r_+ r_- - r_-^2)}{4 \pi r_+^2 L^2} = \frac{(r_+ - r_-) (r_c - r_+) (2 r_+ + r_- + r_c)}{4 \pi r_+^2 L^2},
\nonumber\\
&& S_{BH} = \frac{A_+}4 = \pi r_+^2, \qquad \Phi_H = - \frac{Q}{r_+},
\end{eqnarray}
and the Gibbons-Hawking temperature at the cosmological horizon is
\begin{equation}
T_{GH} = -\frac{f'(r_c)}{4 \pi} = \frac{(r_c - r_+) (r_+^2 + 2 r_+ r_c + 3 r_c^2 - L^2)}{4 \pi r_c^2 L^2} = \frac{(r_c - r_+) (r_c - r_-) (2 r_c + r_+ + r_-)}{4 \pi r_c^2 L^2}.
\end{equation}
The ratio of the Gibbons-Hawking temperature to the Hawking temperature is given by
\begin{eqnarray}
\frac{T_{GH}}{T_H} = \frac{1 - r_-/r_c}{1 - r_-/r_+} \times \frac{1 + (r_- + r_+ + r_c)/r_c}{1 + (r_- + r_+ + r_c)/r_+}.
\end{eqnarray}
Note that both the Hawking temperature and Gibbons-Hawking temperature vanish in the extremal limit of $r_+ = r_c$. Otherwise, they are equal, called the ``lukewarm'' limit, $T_H = T_{GH} = (r_c - r_+)/2 \pi L^2$ when $r_c = r_- (r_+ + r_-)/(r_+ - r_-)$, equivalently, $M = Q$.

\begin{figure}
\includegraphics[scale=0.4, angle=0]{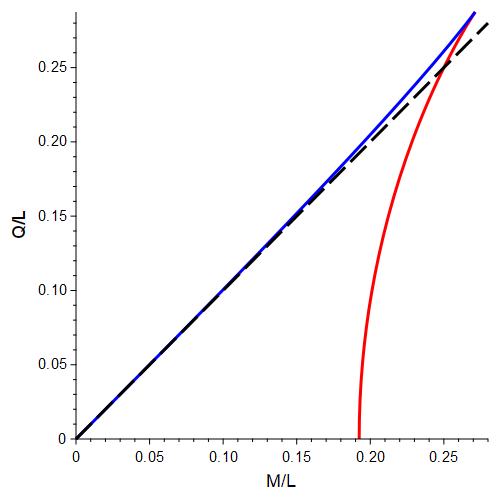}
\caption{Two extremal limits of charged RN-dS black holes: (i) the upper (blue) curve is the extremal RN black hole with the geometry $\mathrm{AdS}_2 \times \mathrm{S}^2$ and (ii) the lower (red) curve is the Nariai black hole with the geometry $\mathrm{dS}_2 \times \mathrm{S}^2$. The coincidence limit of the three horizons $r_- = r_+ = r_c$ is the ultracold black hole. The dashed line $Q = M$  is the lukewarm limit. The region enclosed by the blue and red curves corresponds to nonextremal black holes while the outside region corresponds to spacetimes with naked singularity.}
\label{fig_QvsM}
\end{figure}

In the parameter space of the RN-dS black holes, as shown in Fig.~\ref{fig_QvsM}, the blue curve ($r_+ = r_-$) and the red one ($r_+ = r_c$) divide the parameter space into the two regions: the inside region for RN-dS black holes with three roots and the outside region for black holes with only one root and a naked singularity. In order for the pair production not to drive the Nariai black holes to those with naked singularity, the black holes should lose more mass than charge.
The slope $dM/dQ$ for the Nariai black holes monotonically increases with respect to $Q$ and reaches the maximal value at the ultracold point, $dM/dQ|_\mathrm{max} = \sqrt2$. Therefore, the ``sufficient'' condition to avoid the formation of a singular spacetime is, for the mass and charge of an emitted particle,
\begin{equation} \label{eq_S_cond}
m > \sqrt2 q.
\end{equation}

A charged RN black hole in dS space can have two coincidence limits for extremal black holes: $r_- = r_+$ or $r_+ = r_c$, where $M, Q$, and $L$ satisfy the constraint
\begin{eqnarray}
\left( 1 - 18 \frac{3 M^2 - 2 Q^2}{L^2} \right)^2 = \left( 1 - 12 \frac{Q^2}{L^2} \right)^3.
\end{eqnarray}
The corresponding physical properties in near-extremal limits have been discussed in~\cite{Castro:2022cuo}.
The pair production for the near extremal limit $r_- \sim r_+$ has been discussed in detail~\cite{Chen:2020mqs}. Here we will study the pair production in the other extremal limit, namely Nariai limit as $r_+ = r_c = r_n$ and $M = M_n, Q = Q_n$ where
\begin{equation} \label{eq_NariaiLimit}
r_n^2 = \frac{L^2}6 \left( 1 + \sqrt{1 - 12 Q_n^2/L^2} \right), \qquad M_n = \frac{r_n}3 \left( 2 - \sqrt{1 - 12 Q_n^2/L^2} \right).
\end{equation}
The radius $L$ has a minimal value, $L_\mathrm{min} = \sqrt{12} \, Q_n$, corresponding to the ultracold limit $r_- = r_+ = r_c$, and a Schwarzschild-dS black hole has the Nariai limit when $M/L = 1/\sqrt{27}$. We further consider the near Nariai limit (near-extremal Nariai black hole) with a slight derivation from~\eqref{eq_NariaiLimit} as
\begin{equation}
r_+ = r_n - \epsilon B, \qquad r_c = r_n + \epsilon B, \qquad M = M_n - \epsilon^2 B^2 \frac{2 r_n}{L^2}, \qquad Q^2 = Q_n^2 - \epsilon^2 B^2 \frac{M_n}{r_n}.
\end{equation}
Then the Hawking temperature reduces to
\begin{equation}
T_H = \frac{B}{2 \pi} \frac{\epsilon}{r_\mathrm{ds}^2},
\end{equation}
in which an important scale, the radius of dS$_2$ appearing in the near-horizon geometry, is defined as
\begin{equation} \label{eq_rds}
r_\mathrm{ds}^2 = \frac{r_n^2}{6 r_n^2/L^2 - 1} = \frac{L^2}{6} \left( \frac{1}{\sqrt{1 - 12 Q_n^2/L^2}} + 1 \right).
\end{equation}
Note that the geometry of the Nariai black hole has the structure of a product space, $\mathrm{dS}_2 \times \mathrm{S}^2$~\cite{Ortaggio:2002bp},
\begin{equation}
\frac{1}{r_\mathrm{ds}^2} + \frac{1}{r_n^2} = \frac{6}{L^2} = 2 \Lambda,
\end{equation}
in contrast to the geometry of extremal RN black hole, $\mathrm{AdS}_2 \times \mathrm{S}^2$,
\begin{equation}
- \frac{1}{r_\mathrm{ads}^2} + \frac{1}{r_n^2} =  \frac{6}{L^2} = - 2 \Lambda.
\end{equation}

\begin{figure}
\includegraphics[scale=0.6, angle=0]{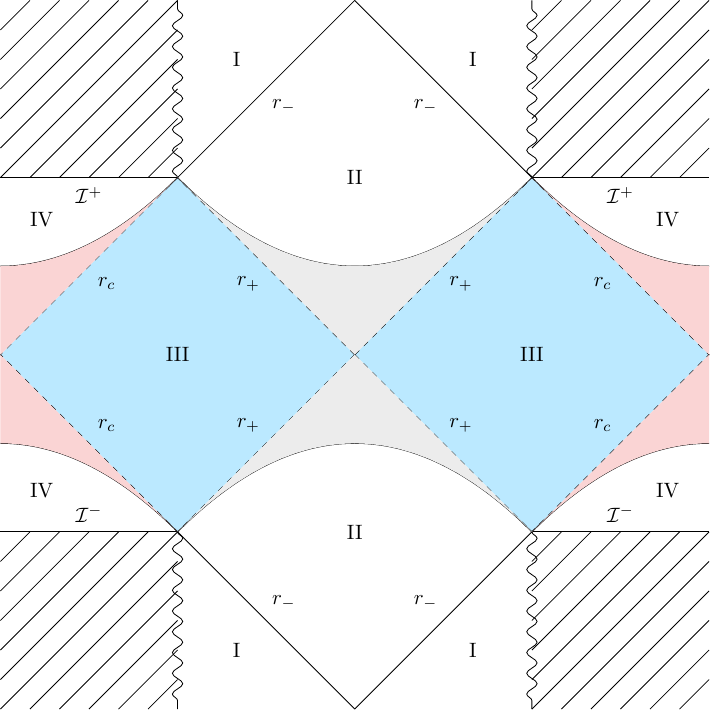}
\caption{The Penrose-Cater diagram of RN-dS black holes. The colored zone denotes the near horizon region of near extremal Nariai black holes with a tiny region III. ${\cal I}$ denotes the infinity $r = \infty$.}
\label{fig_PenrosDiagram}
\end{figure}

In Fig.~\ref{fig_PenrosDiagram} of the Penrose-Carter diagram of RN-dS black holes~\cite{Gibbons:1977mu, Cardoso:2004uz, Costa:2018zvw}, the Killing vector $K = \partial/\partial t$ becomes timelike, future-directed in the region III ($r_+ < r < r_c$) and spacelike in the region IV ($r > r_c$), so we will call these regions ``timelike inner region'' and ``spacelike outer region,'' respectively. Charged pairs are produced at the near horizon region of near-extremal Nariai black holes, namely in the colored zones, specifically narrow regions about III. Provided that the decay of charge destroys both $r_+$ and $r_c$, then the space-like regions II and IV coalesce, and the only remaining time-like region I is causally connected to the singularity which is a singular spacetimes.

\begin{figure}
\includegraphics[scale=0.2, angle=0]{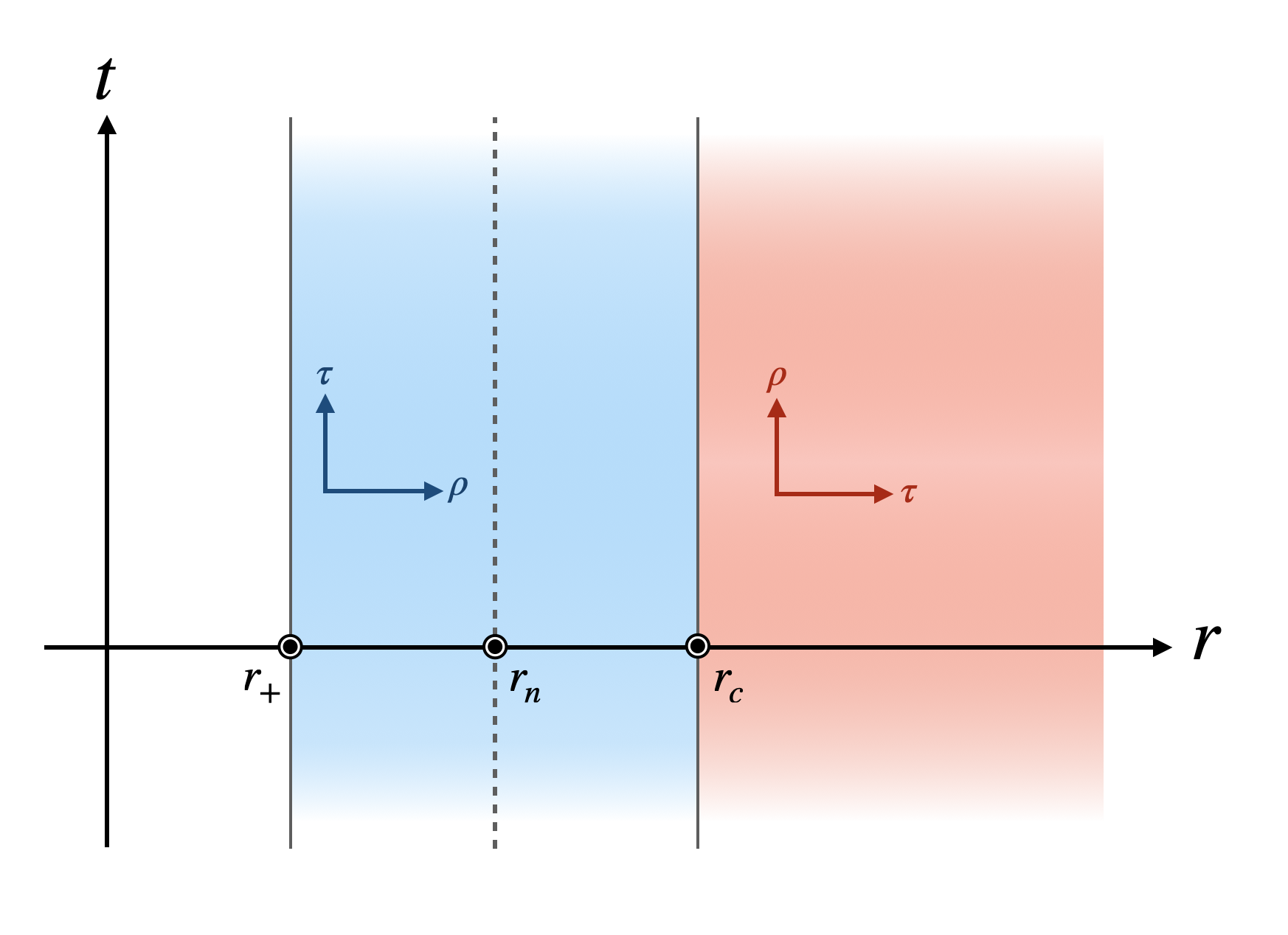}
\caption{The two regions of Nariai black holes are studied: (i) spacelike outer region $r > r_c$ and (ii) timelike inner region $r_+ < r < r_c$}
\label{fig_coordinates}
\end{figure}

The near-extremal Nariai has two interesting regions to study the pair production, see Fig.~\ref{fig_coordinates}: (i) spacelike outer region $r > r_c$ and (ii) timelike inner region $r_+ < r < r_c$, which will be investigated separately below.

\subsection{Spacelike Outer Region}
%%%%%%%%%%%%%%%%%%%%%%%%%%%%%%%%%%%%%%%%%%%%%%%%%%%%%%%%%%%%%%%%%%%%%%
The geometry of the spacelike outer region of near-extremal Nariai black hole can be represented by a suitable coordinates $(\tau, \rho)$ as
\begin{equation}
r = r_n + \epsilon \, \tau, \qquad t = \frac{r_\mathrm{ds}^2}{\epsilon} \, \rho.
\end{equation}
Then by taking $\epsilon \to 0$, one can get the near-horizon geometry ($\tau > B$)
\begin{equation} \label{eq_outNariai}
ds^2 = r_\mathrm{ds}^2 \left[ - \frac{d\tau^2}{\tau^2 - B^2} + (\tau^2 - B^2) d\rho^2 \right] + r_n^2 d\Omega_2^2,
\end{equation}
in which the gauge field (the sign of charge is chosen such that the electric field points the positive $\rho$-direction) is given by
\begin{equation}
A_{[1]} = \frac{r_\mathrm{ds}^2 Q_n}{r_n^2} \tau d\rho, \qquad \qquad F_{[2]} = \frac{r_\mathrm{ds}^2 Q_n}{r_n^2} d\tau \wedge d\rho = \frac{Q_n}{r_n^2} \, \vartheta^\tau \wedge \vartheta^\rho = E \, \vartheta^\tau \wedge \vartheta^\rho,
\end{equation}
where $\vartheta^\tau = r_\mathrm{ds} d\tau/\sqrt{\tau^2 - B^2}, \, \vartheta^\rho = r_\mathrm{ds} \sqrt{\tau^2 - B^2} d\rho$ are the orthonormal frames. The role of $t$ and $r$ is interchanged, and~\eqref{eq_outNariai} describes a time-dependent, expanding geometry with dS$_2 \times S^2$ structure.

\subsection{Timelike Inner Region}
%%%%%%%%%%%%%%%%%%%%%%%%%%%%%%%%%%%%%%%%%%%%%%%%%%%%%%%%%%%%%%%%%%%%%%
For the timelike inner region, it is more convenient to use the following coordinates
\begin{equation}
r = r_n + \epsilon \rho, \qquad t = \frac{r_\mathrm{ds}^2}{\epsilon} \, \tau.
\end{equation}
Then the near-horizon geometry ($-B < \rho < B$) describes a static geometry of $\mathrm{dS}_2 \times \mathrm{S}^2$:
\begin{equation} \label{eq_inNariai}
ds^2 = r_\mathrm{ds}^2 \left[ -(B^2 - \rho^2) d\tau^2 + \frac{d\rho^2}{B^2 - \rho^2} \right] + r_n^2 d\Omega_2^2,
\end{equation}
and the gauge field is
\begin{equation}
A_{[1]} = - \frac{r_\mathrm{ds}^2 Q_n}{r_n^2} \rho d\tau, \qquad F_{[2]} = \frac{r_\mathrm{ds}^2 Q_n}{r_n^2} d\tau \wedge d\rho = \frac{Q_n}{r_n^2} \, \vartheta^\tau \wedge \vartheta^\rho = E \, \vartheta^\tau \wedge \vartheta^\rho.
\end{equation}
Here, the orthonormal frames are $\vartheta^\tau = r_\mathrm{ds} \sqrt{B^2 - \rho^2} d\tau$ and $\vartheta^\rho = r_\mathrm{ds} d\rho/ \sqrt{B^2 - \rho^2}$.

%%%%%%%%%%%%%%%%%%%%%%%%%%%%%%%%%%%%%%%%%%%%%%%%%%%%%%%%%%%%%%%%%%%%%%
\section{Pair Production} \label{sec III}
%%%%%%%%%%%%%%%%%%%%%%%%%%%%%%%%%%%%%%%%%%%%%%%%%%%%%%%%%%%%%%%%%%%%%%
The action for a probe charged scalar field $\Phi$ with mass $m$ and charge $q$ in a curved spacetime is
\begin{equation} \label{action}
S (\phi, \phi^*) = \int d^4x \sqrt{-g} \left( - \frac12 D_\alpha \Phi^* D^\alpha \Phi - \frac12 m^2 \Phi^2 \right),
\end{equation}
where the derivative $D_\alpha$ is defined as $D_{\alpha} \equiv \nabla_{\alpha} - i q A_{\alpha}$, and $\nabla_\alpha$ is the covariant derivative in the spacetime. The corresponding Klein-Gordon (KG) equation is
\begin{equation} \label{eom}
(\nabla_\alpha - i q A_\alpha) (\nabla^\alpha - i q A^\alpha) \Phi - m^2 \Phi = 0.
\end{equation}

\subsection{Spacelike Outer Region}
%%%%%%%%%%%%%%%%%%%%%%%%%%%%%%%%%%%%%%%%%%%%%%%%%%%%%%%%%%%%%%%%%%%%%%
For the outer region, the background spacetime is time-dependent, and pair production, from $\tau = B$ to $\tau = \infty$, is analogous to a scattering process over a time-dependent potential. Using the symmetry of Nariai black hole, the scalar field
\begin{equation} \label{ansatz}
\Phi(\tau, \rho, \theta, \phi) = \mathrm{e}^{i k \rho} T(\tau) Y_l^n(\theta, \phi)
\end{equation}
with the standard spherical harmonics $Y_l^n(\theta, \phi)$, satisfies the mode equation $T(\tau)$ of the KG equation
\begin{equation} \label{eq_KGt}
\frac{d}{d\tau} \left[ (\tau^2 - B^2) \frac{d}{d\tau} T \right] + \left[ \frac{(r_\mathrm{ds}^2 q Q_n \tau - r_n^2 k)^2}{r_n^4 (\tau^2 - B^2)} + r_\mathrm{ds}^2 m^2 + \frac{r_\mathrm{ds}^2}{r_n^2} l (l + 1) \right] T = 0.
\end{equation}
The general solution for the KG equation is given by the Gauss hypergeometric function
\begin{eqnarray}
T(\tau) &=& c_1 (\tau + B)^{i (\tilde{\kappa} + \kappa)/2} (\tau - B)^{-i (\tilde{\kappa} - \kappa)/2} F\left( \frac12 + i \kappa + i \mu, \frac12 + i \kappa - i \mu; 1 - i \tilde{\kappa} + i \kappa; z \right)
\nonumber\\
&+& c_2 (\tau + B)^{i (\tilde{\kappa} + \kappa)/2} (\tau - B)^{i (\tilde{\kappa} - \kappa)/2} F\left( \frac12 + i \tilde{\kappa} + i \mu, \frac12 + i \tilde{\kappa} - i \mu; 1 + i \tilde{\kappa} - i \kappa; z \right),
\end{eqnarray}
where
\begin{equation}
\tilde{\kappa} = \frac{k}{B}, \qquad \kappa = q Q_n \frac{r_\mathrm{ds}^2}{r_n^2}, \qquad \mu^2 = q^2 Q_n^2 \frac{r_\mathrm{ds}^4}{r_n^4} + m^2 r_\mathrm{ds}^2 + l (l + 1) \frac{r_\mathrm{ds}^2}{r_n^2} - \frac14, \qquad z = - \frac{\tau - B}{2 B}.
\end{equation}
The necessary condition for pair production is that the parameter $\mu$ should be real, which gives propagating modes to emitted particles, as is shown in~\eqref{eq_KGt} in the large $\tau$ limit and will be explicitly shown by the powers of $\tau$ in~\eqref{eq_Pmode1} and~\eqref{eq_Pmode2} below. However, this doe not ensure the sufficient condition~\eqref{eq_S_cond}, therefore the pair production may drive Nariai black holes to those with a naked singularity by emitting light charged particle.

We find the in- and out-going modes at the initial time ($\tau = B$) and the final time ($\tau \to \infty$) and compute the associated energy densities by
\begin{equation}
D = i \sqrt{-g} g^{\tau\tau} (\Phi \nabla_\tau \Phi^* - \Phi^* \nabla_\tau \Phi).
\end{equation}
The in- and out-going modes at initial time, $\tau = B$ ($z = 0$), and their associated energy densities are\footnote{The Bogoliubov coefficients depend only on the density ratios. Thus, here and after, a common irrelevant factor from $\sqrt{-g}$ is neglected in each density.}
\begin{eqnarray} \label{eq_PhiBout}
\Phi_B^\rightarrow &=& c_1 (2 B)^{i (\tilde{\kappa} + \kappa)/2} (\tau - B)^{-i (\tilde{\kappa} - \kappa)/2} \qquad \Rightarrow \qquad D_B^\rightarrow = |c_1|^2 (\tilde{\kappa} - \kappa) \frac{2 B}{r_\mathrm{ds}^2},
\\
\Phi_B^\leftarrow &=& c_2 (2 B)^{i (\tilde{\kappa} + \kappa)/2} (\tau - B)^{i (\tilde{\kappa} - \kappa)/2}, \qquad \Rightarrow \qquad D_B^\leftarrow = - |c_2|^2 (\tilde{\kappa} - \kappa) \frac{2 B}{r_\mathrm{ds}^2}.
\end{eqnarray}
The boundary condition to be imposed is $D_B^\leftarrow = 0$, namely $c_2 = 0$, and then the in- and out-going modes at the final time $\tau \to \infty$ are
\begin{eqnarray}
\Phi_\infty^\rightarrow &=& c_1 (2 B)^{1/2 + i \kappa + i \mu} \frac{\Gamma(1 - i \tilde{\kappa} + i \kappa) \Gamma(-i 2 \mu)}{\Gamma(1/2 + i \kappa - i \mu) \Gamma(1/2 - i \tilde{\kappa} - i \mu)} \tau^{-1/2 - i \mu} \label{eq_Pmode1}
\\
&\Rightarrow& D_\infty^\rightarrow = |c_1|^2 (\tilde{\kappa} - \kappa) \frac{2 B}{r_\mathrm{ds}^2} \, \frac{\cosh(\pi \kappa - \pi \mu) \cosh(\pi \tilde{\kappa} + \pi \mu)}{\sinh(\pi \tilde{\kappa} - \pi \kappa) \sinh(2 \pi \mu)},
\\
\Phi_\infty^\leftarrow &=& c_1 (2 B)^{1/2 + i \kappa - i \mu} \frac{\Gamma(1 - i \tilde{\kappa} + i \kappa) \Gamma(i 2 \mu)}{\Gamma(1/2 + i \kappa + i \mu) \Gamma(1/2 - i \tilde{\kappa} + i \mu)} \tau^{-1/2 + i \mu} \label{eq_Pmode2}
\\
&\Rightarrow& D_\infty^\leftarrow = - |c_1|^2 (\tilde{\kappa} - \kappa) \frac{2 B}{r_\mathrm{ds}^2} \, \frac{\cosh(\pi \kappa + \pi \mu) \cosh(\pi \tilde{\kappa} - \pi \mu)}{\sinh(\pi \tilde{\kappa} - \pi \kappa) \sinh(2 \pi \mu)}.
\end{eqnarray}
It is straightforward to check the energy conservation, $D_\infty^\rightarrow + D_\infty^\leftarrow = D_B^\rightarrow$. The problem describes a scattering process, and the mean number for pair production is
\begin{eqnarray} \label{eq_Nout}
&& \mathcal{N}_\mathrm{out} = - \frac{D_\infty^\leftarrow}{D_B^\rightarrow} = \frac{\cosh(\pi \kappa + \pi \mu) \cosh(\pi \tilde{\kappa} - \pi \mu)}{\sinh(\pi \tilde{\kappa} - \pi \kappa) \sinh(2 \pi \mu)}, \qquad \mathrm{for} \quad \tilde{\kappa} \ge \kappa,
\\
\stackrel{B \to 0}{\longrightarrow} && \mathcal{N}_\mathrm{out} = \mathrm{e}^{-\pi (\mu - \kappa)} \frac{\cosh(\pi \kappa + \pi \mu)}{\sinh(2 \pi \mu)}.
\nonumber
\end{eqnarray}
As a passing remark, we note that the mean number (\ref{eq_Nout}) has the same form as that for Schwinger pair production in a pulsed Sauter-type electric field~\cite{Kim:2008yt}.

It is interesting to give a thermodynamic interpretation. For pair production, the values of parameters $\mu, \kappa$ generically are $\mu \sim \kappa \gg 1$, and thus the mean number can naturally be expressed as
\begin{equation}
\mathcal{N}_\mathrm{out} = \frac{\cosh(\pi \kappa + \pi \mu)}{\sinh(2 \pi \mu)} \frac{\cosh(\pi \tilde{\kappa} - \pi \mu)}{\sinh(\pi \tilde{\kappa} - \pi \kappa)} = \mathrm{e}^{-2 \pi(\mu - \kappa)} \frac{1 + \mathrm{e}^{-2 \pi (\mu + \kappa)}}{1 - \mathrm{e}^{-2 \pi (\mu - \kappa)} \mathrm{e}^{-2 \pi (\mu + \kappa)}} \frac{1 + \mathrm{e}^{-2 \pi (\tilde{\kappa} - \kappa)} \mathrm{e}^{2 \pi (\mu - \kappa)}}{1 - \mathrm{e}^{-2 \pi (\tilde{\kappa} - \kappa)}}.
\end{equation}
The parameter $\kappa$ is related to the Unruh temperature for charge acceleration by the electric force $F = q E = q Q_n/r_n^2$
\begin{equation}
\kappa = F r_\mathrm{ds}^2, \qquad 2 \pi T_U = F / \bar{m} \quad \Rightarrow \quad \kappa = 2 \pi T_U \bar{m} r_\mathrm{ds}^2.
\end{equation}
By introducing an ``effective inertial mass'' $\bar{m}$ as
\begin{equation}
\mu^2 - \kappa^2 = m^2 r_\mathrm{ds}^2 + l (l +1) \frac{r_\mathrm{ds}^2}{r_n^2} - \frac14 = \bar{m}^2 r_\mathrm{ds}^2,
\end{equation}
the exponents can be rewritten as
\begin{eqnarray}
2 \pi (\mu + \kappa) &=& \frac{\mu^2 - \kappa^2}{(\mu - \kappa)/2 \pi} = \frac{\bar{m}^2 r_\mathrm{ds}^2}{\left( \sqrt{\kappa^2 + \bar{m}^2 r_\mathrm{ds}^2} - \kappa \right)/2 \pi} = \frac{\bar m}{\sqrt{T_U^2 + T_C^2} - T_U},
\nonumber\\
2 \pi (\mu - \kappa) &=& \frac{\mu^2 - \kappa^2}{(\mu + \kappa)/2 \pi} = \frac{\bar{m}^2 r_\mathrm{ds}^2}{\left( \sqrt{\kappa^2 + \bar{m}^2 r_\mathrm{ds}^2} + \kappa \right)/2 \pi} = \frac{\bar m}{\sqrt{T_U^2 + T_C^2} + T_U},
\\
2 \pi (\tilde{\kappa} - \kappa) &=& \frac{2 \pi k}{B} - \frac{2 \pi q Q_n r_\mathrm{ds}^2}{r_n^2} = \frac{k - q \Phi_H}{T_H},
\nonumber
\end{eqnarray}
where $T_C$ is the temperature associated to the dS$_2$ curvature, and $T_H, \Phi_H$ are the Hawking temperature (in rescaled coordinates) and chemical potential, respectively,
\begin{equation}
T_C = \frac1{2 \pi r_\mathrm{ds}}, \qquad T_H = \frac{B}{2 \pi}, \qquad \Phi_H = \frac{Q_n B r_\mathrm{ds}^2}{r_n^2}.
\end{equation}
Finally, we find the mean number in terms of the effective temperatures as the universal form
\begin{equation}
\mathcal{N}_\mathrm{out} = \Biggl( \frac{1 + \mathrm{e}^{-\bar{m}/\bar{T}_\mathrm{eff}}}{1 - \mathrm{e}^{-\bar{m}/T_\mathrm{eff}} \mathrm{e}^{-\bar{m}/{\bar T}_\mathrm{eff}}} \Biggr) \times  \mathrm{e}^{-\bar{m}/T_\mathrm{eff}} \times \Biggl( \frac{1 + \mathrm{e}^{-(k - q \Phi_H)/T_H} \mathrm{e}^{\bar{m}/T_\mathrm{eff}}}{1 - \mathrm{e}^{-(k - q \Phi_H)/T_H} } \Biggr),
\end{equation}
where
\begin{equation} \label{eff-tem}
T_\mathrm{eff} = \sqrt{T_U^2 + T_C^2} + T_U, \qquad {\bar T}_\mathrm{eff} = \sqrt{T_U^2 + T_C^2} - T_U.
\end{equation}
The dominant term for the charge emission is the Boltzmann factor $\mathrm{e}^{-\bar{m}/T_\mathrm{eff}}$ because $T_\mathrm{eff} \gg \bar{T}_\mathrm{eff}$, which is satisfied by $E r_{\rm ds} \gg \bar{m}/e$ in the standard QED except for small black holes. Remarkably, there is a bosonlike condensation when $| k - q \Phi_H |  \ll T_H$, which catastrophically explodes for $k = q \Phi_H$.

For Nariai black hole, we take $B = 0 \; (T_H = 0)$ limit and obtain
\begin{eqnarray} \label{Nariai-emission}
\mathcal{N}_\mathrm{out} = \frac{\mathrm{e}^{-2 \pi (\mu - \kappa)} + \mathrm{e}^{-4 \pi \mu} }{1 - \mathrm{e}^{-4 \pi \mu}}
= \frac{\mathrm{e}^{-\bar{m}/T_\mathrm{eff}} + \mathrm{e}^{-\bar{m}/[ T_C /( 2 \sqrt{1 + T_U^2/T_C^2} ) ]}}{1 - \mathrm{e}^{-\bar{m}/[ T_C /( 2 \sqrt{1 + T_U^2/T_C^2} ) ]}}.
\end{eqnarray}
The emission formula~\eqref{Nariai-emission} for a charge in the S-wave is identical to the Schwinger formula in the planar coordinates of $\mathrm{dS}_2$ space~\cite{Cai:2014qba} by identifying $E = Q_n/r_n^2$ and $H = 1/r_\mathrm{ds}$. The out-vacuum in~\cite{Cai:2014qba} is the asymptotic future limit, where the wavelength is infinitely red-shifted, and the mean number depends only on $\mu$ and $\kappa$.

For the purpose of computing the mean number, the monodromy method~\cite{Chen:2022hpe}, which is briefly summarized in Appendix~\ref{app_B}, indeed provides one with a general formula that straightforwardly gives the final result. By comparing~\eqref{eq_KGt} with the standard Riemann differential equation~\eqref{eq_Riemann}, one readily writes down the solution as the P-function~\eqref{eq_Pfun}
\begin{equation} \label{eq_Pout}
T(\tau) = P\begin{pmatrix} -B & B & \infty & \\ -i (\tilde{\kappa} + \kappa)/2 & -i (\tilde{\kappa} - \kappa)/2 & 1/2 - i \mu & ; \tau \\ i (\tilde{\kappa} + \kappa)/2 & i (\tilde{\kappa} - \kappa)/2 & 1/2 + i \mu & \end{pmatrix}.
\end{equation}
Using~\eqref{eq_Ns} we simply compute the mean number for pair production, as a scattering process, from $\tau = B$ to $\tau \to \infty$. It exactly gives the result~\eqref{eq_Nout} since $\alpha_\infty = 1/2 - i \mu$ and $\beta_\infty = 1/2 + i \mu$ in~\eqref{eq_Ns} change sine functions in the numerator into cosine functions.

\begin{figure}
\includegraphics[scale=0.5, angle=0]{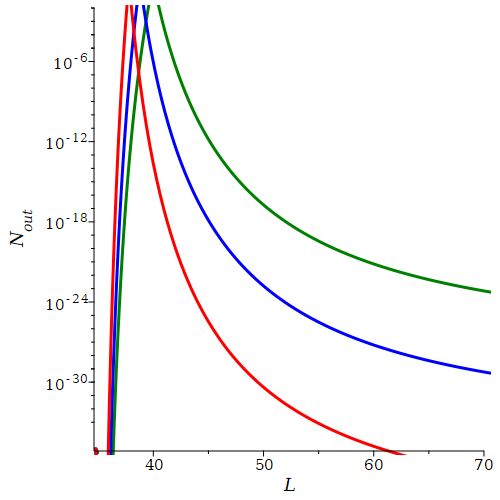}
\hspace{.5cm}
\includegraphics[scale=0.5, angle=0]{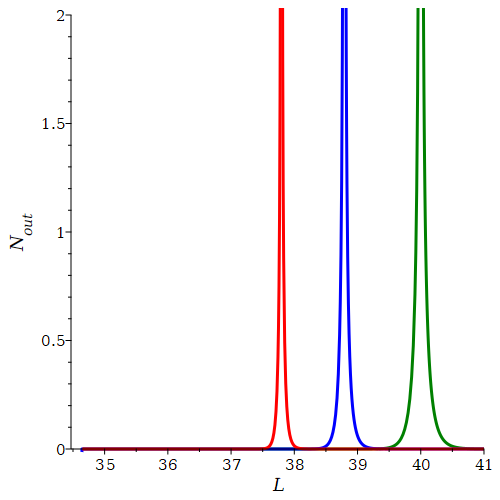}
\caption{The mean number of the pair production in the spacelike outer region: [left] $\mathcal{N}_\mathrm{out}$ (log scale) with respect to $L$, [right] $\mathcal{N}_\mathrm{out}$ for $L$ near the critical value, with parameters $l = 0, m = q = 1, k = 2, Q_n = 10$ and $B = 0.08$ (red), $0.09$ (blue) and $0.1$ (green). The critical values are $L_\mathrm{cr} = 37.8 \, (B = 0.08), \; L_\mathrm{cr} = 38.79 \, (B = 0.09), \; L_\mathrm{cr} = 40 \, (B = 0.1)$ and $L_\mathrm{min} = 34.64$.}
\label{fig_Nout}
\end{figure}

The mean number~\eqref{eq_Nout} is valid with positive value for $\tilde{\kappa} \ge \kappa$ or $k \ge q \Phi_H$, otherwise the classification of in- and out-modes are not correct. Therefore, $\tilde{\kappa} = \kappa$ determines a critical value of $L$
\begin{equation} \label{eq_Lcr}
L_\mathrm{cr}^2 = \frac{12 \omega^2 Q_n^2}{\omega^2 - q^2 Q_n^2 B^2},
\end{equation}
which is greater than the lower bound $L_\mathrm{cr} > L_\mathrm{min} = \sqrt{12} \, Q_n$, i.e. the ultracold limit, see Fig.~\ref{fig_Nout}. For the case $\tilde{\kappa} < \kappa$, the in- and out-modes~\eqref{eq_PhiBout} at $\tau = B$ interchange, and the associated P-function is~\eqref{eq_Pout} with exchange of the characteristic exponents at $\tau = B$. It is equivalent to $\tilde{\kappa} \leftrightarrow \kappa$, therefore the mean number can be straightforwardly obtained from~\eqref{eq_Nout}
\begin{equation}
\mathcal{N}_\mathrm{out} = \frac{\cosh(\pi \tilde\kappa + \pi \mu) \cosh(\pi \kappa - \pi \mu)}{\sinh(\pi \kappa - \pi \tilde{\kappa}) \sinh(2 \pi \mu)}, \qquad \mathrm{for} \quad \tilde\kappa < \kappa.
\end{equation}

\subsection{Timelike Inner Region}
%%%%%%%%%%%%%%%%%%%%%%%%%%%%%%%%%%%%%%%%%%%%%%%%%%%%%%%%%%%%%%%%%%%%%%
In the timelike inner region, the background spacetime is static, and pair production, from $\rho = -B$ to $\rho = B$, is analogous to a tunneling process through a potential barrier. The scalar field is decomposed into the spherical harmonic and a positive frequency mode
\begin{equation} \label{ansatzInt}
\Phi(\tau, \rho, \theta, \phi) = \mathrm{e}^{-i \omega \tau} R(\rho) Y_l^n(\theta, \phi),
\end{equation}
and then the radial mode of the KG equation for $R(\rho)$ reduces to
\begin{equation}
\frac{d}{d\rho} \left[ (B^2 - \rho^2) \frac{d}{d\rho} R \right] + \left[ \frac{(r_\mathrm{ds}^2 q Q_n \rho - r_n^2 \omega)^2}{r_n^4 (B^2 - \rho^2)} - r_\mathrm{ds}^2 m^2 - \frac{r_\mathrm{ds}^2}{r_n^2} l (l + 1) \right] R = 0.
\end{equation}
The general solution is given again by the Gauss hypergeometric function
\begin{eqnarray}
R(\rho) &=& c_1 (B + \rho)^{-i (\tilde{\kappa} + \kappa)/2} (B - \rho)^{-i (\tilde{\kappa} - \kappa)/2} F\left( \frac12 - i \tilde{\kappa} + i \mu, \frac12 - i \tilde{\kappa} - i \mu; 1 - i \tilde{\kappa} + i \kappa; z \right)
\nonumber\\
&+& c_2 (B + \rho)^{-i (\tilde{\kappa} + \kappa)/2} (B - \rho)^{i (\tilde{\kappa} - \kappa)/2} F\left( \frac12 - i \kappa + i \mu, \frac12 - i \kappa - i \mu; 1 + i \tilde{\kappa} - i \kappa; z \right),
\end{eqnarray}
where
\begin{equation}
\tilde{\kappa} = \frac{\omega}{B}, \qquad \kappa = q Q_n \frac{r_\mathrm{ds}^2}{r_n^2}, \qquad \mu^2 = q^2 Q_n^2 \frac{r_\mathrm{ds}^4}{r_n^4} + m^2 r_\mathrm{ds}^2 + l (l +1) \frac{r_\mathrm{ds}^2}{r_n^2} - \frac14, \qquad z = - \frac{\rho - B}{2 B}.
\end{equation}

To define the in- and out-vacua, we decompose the general solution into the in- and out-going modes at $-B$ and $B$ according to the fluxes
\begin{equation}
D = i \sqrt{-g} g^{\rho\rho} (\Phi \nabla_\rho \Phi^* - \Phi^* \nabla_\rho \Phi).
\end{equation}
First, we obtain the in- and out-going modes and derive their fluxes at boundary $\rho = B$ ($z = 0$), i.e. the cosmological horizon
\begin{eqnarray}
\Phi_B^\rightarrow &=& c_1 (2 B)^{-i (\tilde{\kappa} + \kappa)/2} (B - \rho)^{-i (\tilde{\kappa} - \kappa)/2} \qquad \Rightarrow \qquad D_B^\rightarrow = |c_1|^2 (\tilde{\kappa} - \kappa) \frac{2 B}{r_\mathrm{ds}^2},
\\
\Phi_B^\leftarrow &=& c_2 (2 B)^{-i (\tilde{\kappa} + \kappa)/2} (B - \rho)^{i (\tilde{\kappa} - \kappa)/2} \qquad \Rightarrow \qquad D_B^\leftarrow = - |c_2|^2 (\tilde{\kappa} - \kappa) \frac{2 B}{r_\mathrm{ds}^2}.
\end{eqnarray}
According to~\cite{Chen:2012zn}, we impose the boundary condition $D_B^\leftarrow = 0$, i.e. $c_2 = 0$, which corresponds to the zero in-going flux at the cosmological horizon. Then the fluxes at the other boundary $\rho = - B$ (outer horizon of black holes) are
\begin{eqnarray}
\Phi_{-B}^\rightarrow &=& c_1 (2 B)^{-i (3 \tilde{\kappa} + \kappa)/2} \frac{\Gamma(1 - i \tilde{\kappa} + i \kappa) \Gamma(-i \tilde{\kappa} - i \kappa)}{\Gamma(1/2 - i \tilde{\kappa} + i \mu) \Gamma(1/2 - i \tilde{\kappa} - i \mu)} (B + \rho)^{i (\tilde{\kappa} + \kappa)/2}
\\
&\Rightarrow& D_{-B}^\rightarrow = |c_1|^2 (\tilde{\kappa} - \kappa) \frac{2 B}{r_\mathrm{ds}^2} \, \frac{\cosh(\pi \tilde{\kappa} + \pi \mu) \cosh(\pi \tilde{\kappa} - \pi \mu)}{\sinh(\pi \tilde{\kappa} + \pi \kappa) \sinh(\pi \tilde{\kappa} - \pi \kappa)},
\\
\Phi_{-B}^\leftarrow &=& c_1 (2 B)^{-i (\tilde{\kappa} - \kappa)/2} \frac{\Gamma(1 - i \tilde{\kappa} + i \kappa) \Gamma(i \tilde{\kappa} + i \kappa)}{\Gamma(1/2 + i \kappa - i \mu) \Gamma(1/2 + i \kappa + i \mu)} (B + \rho)^{-i (\tilde{\kappa} + \kappa)/2}
\\
&\Rightarrow& D_{-B}^\leftarrow = - |c_1|^2 (\tilde{\kappa} - \kappa) \frac{2 B}{r_\mathrm{ds}^2} \, \frac{\cosh(\pi \kappa + \pi \mu) \cosh(\pi \kappa - \pi \mu)}{\sinh(\pi \tilde{\kappa} + \pi \kappa) \sinh(\pi \tilde{\kappa} - \pi \kappa)}.
\end{eqnarray}
The flux conservation, $D_{-B}^\rightarrow + D_{-B}^\leftarrow = D_B^\rightarrow$, holds. The problem describes a tunneling process, and the mean number of pair production is
\begin{equation} \label{eq_Nin}
\mathcal{N}_\mathrm{in} = - \frac{D_B^\rightarrow}{D_{-B}^\leftarrow} = \frac{\sinh(\pi \tilde{\kappa} + \pi \kappa) \sinh(\pi \tilde{\kappa} - \pi \kappa)}{\cosh(\pi \kappa + \pi \mu) \cosh(\pi \mu- \pi \kappa)}, \qquad \mathrm{for} \quad \tilde{\kappa} \ge \kappa.
\end{equation}

It is interesting to note that the mean number~\eqref{eq_Nin} has the same form as that for Schwinger pair production in a localized Sauter-type electric field~\cite{Kim:2009pg}. There is an amplification factor $\exp(2 \pi \tilde{\kappa} - 2 \pi \mu)$. In the limit $B \to 0$, i.e. $\tilde{\kappa} \to \infty$, the tunneling region shrinks to a spherical surface of zero volume and thus the barrier disappears, making the tunneling ``trivial.'' Consequently, the mean number diverges, and the charge emission becomes catastrophic
\begin{equation}
\lim_{\tilde{\kappa} \to \infty} \mathcal{N}_\mathrm{in} = \mathrm{e}^{2 \pi \tilde{\kappa}} \to \infty.
\end{equation}
However, when pairs are catastrophically produced, the back-reaction of radiations cannot be simply neglected. In the in-out formalism, the one-loop effective, complex action from the scattering amplitude, $\langle {\rm out} \vert {\rm in} \rangle = \exp\left( i \int d^4x \sqrt{-g} {\cal L}^{(1)}_{\rm eff} \right)$, which is equivalent to integrating out the path integral for $ S(\phi, \phi^*) $ over $\phi$ and $\phi^*$, gives the back-reaction to the Maxwell theory, and twice the imaginary part is the vacuum persistence amplitude~\cite{Kim:2008yt, Kim:2009pg}. The induced current or energy-momentum tensor, for instance in Refs.~\cite{Frob:2014zka, Kobayashi:2014zza, Bavarsad:2016cxh, Stahl:2016geq, Banyeres:2018aax, Meimanat:2023hjq}, may be used to quantify the back-reaction of radiation in comparison to the classical counter part. Also, when the density of charged particles from the black hole horizon and antiparticles from the cosmological horizon is high enough to allow scatterings, the particles and antiparticles annihilate into radiation of photons. Thus, the Schwinger pair production leads not only to the effective energy-momentum tensor at the one-loop level but subsequent scatterings of pairs, which go beyond of the scope of this paper and will be investigated in detail in a future work.

In fact, the thermal interpretation for the mean number
\begin{eqnarray}
\mathcal{N}_\mathrm{in} &=& \mathrm{e}^{2 \pi (\tilde{\kappa} - \kappa)} \mathrm{e}^{-2 \pi(\mu - \kappa)} \frac{(1 - \mathrm{e}^{-2 \pi (\tilde{\kappa} + \kappa)}) (1 - \mathrm{e}^{-2 \pi (\tilde{\kappa} - \kappa)})}{(1 + \mathrm{e}^{-2 \pi (\mu + \kappa)}) (1 + \mathrm{e}^{-2 \pi (\mu - \kappa)})}
\nonumber\\
&=& \frac{\left( \mathrm{e}^{(\omega - q \Phi_H)/T_H} - 1 \right) \left( 1 - \mathrm{e}^{-(\omega + q \Phi_H)/T_H} \right)}{\left( \mathrm{e}^{\bar{m}/T_\mathrm{eff}} + 1 \right) \left( 1 + \mathrm{e}^{-\bar{m}/\bar{T}_\mathrm{eff}} \right)},
\end{eqnarray}
implies that the bosonic amplification factor $\exp\left[(\omega - q \Phi_H)/T_H - \bar{m}/T_\mathrm{eff} \right]$ results in exponentially large production of charges with high energy, i.e. $\omega \gg q \Phi_H + \bar{m}T_H/T_\mathrm{eff}$. In other words, the emission of charges becomes catastrophic, $\exp[2 \pi (\tilde{\kappa} - \mu)] \gg 1$, provided that
\begin{eqnarray} \label{catast_con}
\frac{\omega}{q} \gg \sqrt{\Bigl( \frac{Q_n r_\mathrm{ds}^2}{r_n^2} \Bigr)^2 + \Bigl( \frac{\bar{m}r_\mathrm{ds}}{q} \Bigr)^2} \, B.
\end{eqnarray}
Thus, charge emission exponentially explodes as the distance between two horizons draws closer and closer. Furthermore, large Nariai black holes with charge $Q_n > \sqrt{2}/B$ evolve to nonextremal RN-dS black holes since~\eqref{catast_con} satisfies \eqref{eq_S_cond}, the condition for no formation of singular spacetime.

Similarly, the mean number can be found from the general formula derived by the monodromy method. The associated P-function in this case is
\begin{equation}
R(\rho) = P\begin{pmatrix} \infty & -B & B & \\ 1/2 - i\mu & -i (\tilde{\kappa} + \kappa)/2 & -i (\tilde{\kappa} - \kappa)/2 & ; \rho \\ 1/2 + i \mu & i (\tilde{\kappa} + \kappa)/2 & i (\tilde{\kappa} - \kappa)/2 & \end{pmatrix}.
\end{equation}
Using the formula~\eqref{eq_Nt}, the mean number for pair production, as tunneling process from $-B$ to $B$, is exactly equal to the result~\eqref{eq_Nin}. Note again that $\alpha_1 = 1/2 - i \mu$ and $\beta_1 = 1/2 + i \mu$ in~\eqref{eq_Nt} change sine functions in the denominator into cosine functions.

\begin{figure}
\includegraphics[scale=0.5, angle=0]{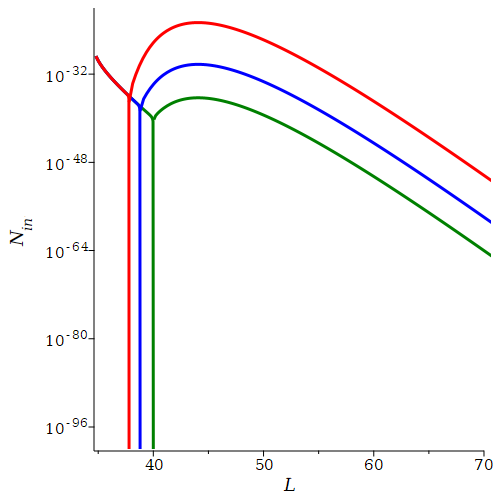}
\hspace{.5cm}
\includegraphics[scale=0.5, angle=0]{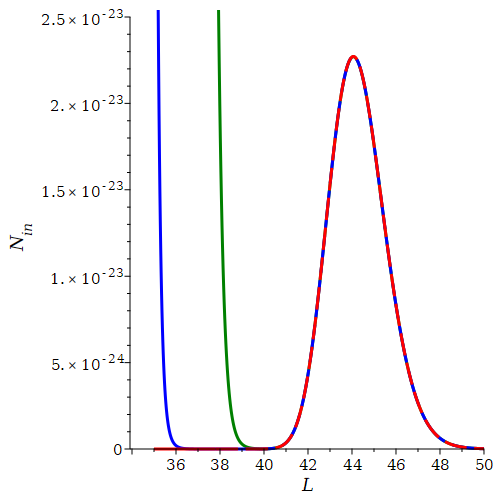}
\caption{The mean number of pair production in the timelike inner region: [left] $\mathcal{N}_\mathrm{in}$ (log scale) with respect to $L$, [right] $\mathcal{N}_\mathrm{in}$ for $L$ near the critical value, with parameters $l = 0, m = q = 1, \omega = 2, Q_n = 10$ and $B = 0.08$ (red), $0.09$ (blue) and $0.1$ (green). In the right panel, their amplitudes are many order different, and therefore suitable amplifications are adapted, i.e. $4.4 \times 10^{13}$ for green line and $3.8 \times 10^7$ for blue line. The critical values are $L_\mathrm{cr} = 37.8 \, (B = 0.08), \; L_\mathrm{cr} = 38.79 \, (B = 0.09), \; L_\mathrm{cr} = 40 \, (B = 0.1)$ and $L_\mathrm{min} = 34.64$.}
\label{fig_Nin}
\end{figure}

\begin{figure}
\includegraphics[scale=0.5, angle=0]{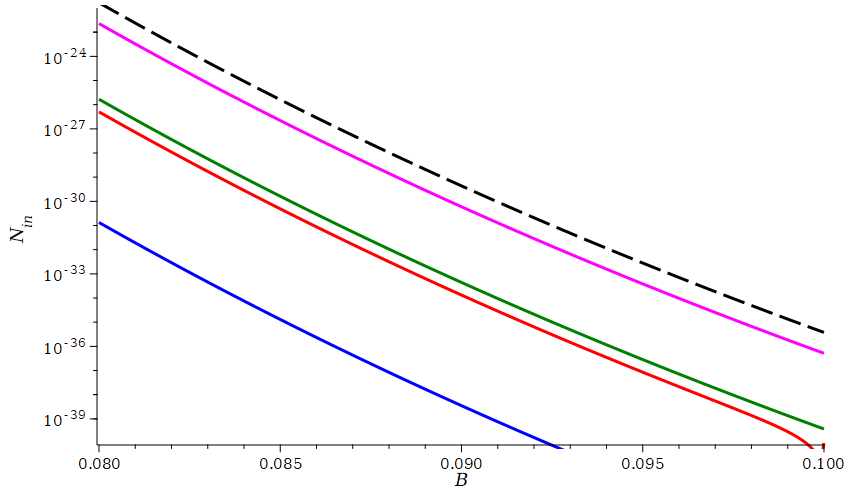}
\caption{The mean number (log scale) of the pair production in the timelike inner region versus to $B$ with parameters $l = 0, m = q = 1, \omega = 2, Q_n = 10$ and $L = 40$ (red), $44$ (magenta) approximately the maxima corresponding to the turning point of $\mathcal{N}_\mathrm{in}$ in Fig.~\ref{fig_Nin}, $50$ (green) and $55$ (blue). All the curves, up to a scale, match with the black dash curve which is $\mathcal{N}_\mathrm{in} = \exp(2 \pi \omega/B)$ minified by a scale $10^{-90}$.}
\label{fig_NinB}
\end{figure}

The mean number~\eqref{eq_Nin} is valid for $\tilde{\kappa} \ge \kappa$, $\omega \ge q \Phi_H$ or $L > L_\mathrm{cr}$. The mean number for $\tilde{\kappa} < \kappa$ can be obtained again from~\eqref{eq_Nin} by exchanging $\tilde{\kappa}$ and $\kappa$
\begin{equation} \label{eq_Nina}
\mathcal{N}_\mathrm{in} = \frac{\sinh(\pi \tilde{\kappa} + \pi \kappa) \sinh(\pi \kappa - \pi \tilde{\kappa})}{\cosh(\pi \tilde{\kappa} + \pi \mu) \cosh(\pi \tilde{\kappa} - \pi \mu)}, \qquad \mathrm{for} \quad \tilde\kappa < \kappa.
\end{equation}

A few comments are in order.
\begin{itemize}
\item The near-extremal Nariai black hole plays an analog of a spherical conductor that breaks down and discharges. The emission of charge $q$ from the black hole horizon is the same as that of $-q$ from the cosmological horizon as shown in the invariance of~\eqref{eq_Nin} under $\kappa$ to $-\kappa$. In the discharging conductor, the current flows from the positive potential to the negative potential.

\item The mean number with respect to $L$ for different $B$ in Fig.~\ref{fig_Nin} is ``universal'' up to a scale when $L$ is sufficiently greater than the critical value $L_\mathrm{cr}$. In other words, when $\tilde\kappa$ is sufficiently greater than $\kappa$ such that $\cosh(2 \pi \tilde\kappa) \gg \cosh(2 \pi \kappa)$, then the mean number~\eqref{eq_Nin} reduces to (generally $\mu > \kappa$)
    \begin{equation} \label{eq_Nin1}
    \mathcal{N}_\mathrm{in} = \frac{\cosh(2 \pi \tilde{\kappa}) - \cosh(2 \pi \kappa)}{\cosh(2 \pi \mu) + \cosh(2 \pi \kappa)} \approx \frac{\cosh(2 \pi \tilde{\kappa})}{\cosh(2 \pi \mu)}, \qquad \mathrm{for} \quad \tilde{\kappa} \gg \kappa.
    \end{equation}
    Thus $1/\cosh(2 \pi \mu)$ determines the profile and $\cosh(2 \pi \tilde{\kappa})$ is $L$-independent ``scale'' which, with $\omega = 2$, leads to
    $$ \frac{\cosh(4 \pi/0.08)}{\cosh(4 \pi/0.09)} = 3.8 \times 10^7, \qquad \frac{\cosh(4 \pi/0.08)}{\cosh(4 \pi/0.1)} = 4.4 \times 10^{13}. $$
    As shown in Fig.~\ref{fig_Nin}, such universality still works well for large $L$ (already close to the critical value). For the case $L$ is sufficiently smaller than $L_\mathrm{cr}$, the mean number~\eqref{eq_Nina} reduces
    \begin{equation} \label{eq_Nin2}
    \mathcal{N}_\mathrm{in} = \frac{\cosh(2 \pi \kappa) - \cosh(2 \pi \tilde{\kappa})}{\cosh(2 \pi \mu) + \cosh(2 \pi \tilde{\kappa})} \approx \frac{\cosh(2 \pi \kappa)}{\cosh(2 \pi \mu)}, \qquad \mathrm{for} \quad \tilde{\kappa} \ll \kappa,
    \end{equation}
    which is independent on $B$.

\item The mean number shown in Fig.~\ref{fig_Nin} has a turning point corresponding to ``local'' maximum of profile $1/\cosh(2 \pi \mu)$. In fact, the turning point agrees with the minimal value of $\mu$ since $\cosh(2 \pi \mu)$ is a monotonically increasing function for positive $\mu$. Therefore, the turning point can be derived by $\partial_L \mu = 0$. For the parameters in Fig.~\ref{fig_Nin} it can be solved numerically $L_\mathrm{turn} = 44.064$ which is consistent with the plots. In more detail, there are two competing contributions in $\mu$, namely electric force (monotonically decreasing) and effective mass (monotonically increasing)
    $$ \mu^2 = F^2 r_\mathrm{ds}^4 + \bar{m}^2 r_\mathrm{ds}^2. $$
    For $L > L_\mathrm{turn}$ the effective mass term dominates, and then $\mu$ is an increasing function leading to decreasing mean number profile. On the other hand, for $L < L_\mathrm{turn}$ the electric force term dominates, and $\mu$ becomes a decreasing function implying increasing mean number profile.

\item In the previous approximation~\eqref{eq_Nin1}, the $B$-dependence of the mean number is about
    \begin{equation}
     \mathcal{N}_\mathrm{in} \approx \cosh(2 \pi \tilde{\kappa}) \approx \mathrm{e}^{2 \pi \omega/B}, \qquad \ln\mathcal{N}_\mathrm{in} \approx \frac{2 \pi \omega}{B}.
    \end{equation}
    As shown in Fig.~\ref{fig_NinB}, all the curves of mean number with different values of $L$, up to a scale, match with this approximation. During the charge emission, for a fixed $L$, $B$ decreases, and hence the mean number exponentially increases and accelerates the discharge process.

\item The pair production in the timelike inner region is a catastrophic process for charge with energy~\eqref{catast_con}; mathematically, the factor $\exp(2 \pi \tilde\kappa - 2 \pi \kappa) = \exp[(\omega - q \Phi_H)/T_H]$ in~\eqref{eq_Nin} has the positive sign opposite to usual Boltzmann factors. The mean number exponentially increases when the temperature proportional to $B$ decreases during the emission. The exponential explosion of charges for $\omega > q \Phi_H$ contrasts with the super-radiant regime for $\omega < q \Phi_H$ in a charged black hole~\cite{Moss:2023kah}. In the charged black hole, $\omega > q \Phi_H$ corresponds to the non-super-radiant regime. Needless to say, the presence of cosmological horizon increases the effective temperature~(\ref{eff-tem}) for charge emission which holds for Nariai black hole with zero Hawking temperature.

\end{itemize}

%%%%%%%%%%%%%%%%%%%%%%%%%%%%%%%%%%%%%%%%%%%%%%%%%%%%%%%%%%%%%%%%%%%%%%
\section{Near-extremal Nariai Black Hole vs RN-dS Black Hole} \label{sec IV}
%%%%%%%%%%%%%%%%%%%%%%%%%%%%%%%%%%%%%%%%%%%%%%%%%%%%%%%%%%%%%%%%%%%%%%
A Nariai black hole has the near-horizon geometry of $\mathrm{dS}_2 \times \mathrm{S}^2$ whereas a near-extremal RN-dS black hole has another near-horizon geometry of $\mathrm{AdS}_2 \times \mathrm{S}^2$. It is illuminating the similarity and difference to compare the emission from the Nariai black hole and the near-extremal RN-dS black hole.

The near-horizon geometry of near-extremal RN-dS black hole is given by~\cite{Chen:2020mqs}
\begin{eqnarray}
ds^2 = - \frac{\rho^2 - B^2}{r_{\mathrm{ads}}^2} d \tau^2 + \frac{r_{\mathrm{ads}}^2}{\rho^2 - B^2} d\rho^2 + r_0^2 d \Omega_2^2,
\end{eqnarray}
where the radius of AdS and the black hole radius are
\begin{eqnarray} \label{RN-radii}
r_\mathrm{ads}^2 = \frac{r_0^2}{1 - 6 r_0^2/L^2} = \frac{L^2}{6} \left( \frac{1}{\sqrt{1 - 12 Q_0^2/L^2}} - 1 \right), \qquad r_0^2 = \frac{L^2}{6} \left( 1 - \sqrt{1 - 12 Q_0^2/L^2} \right).
\end{eqnarray}
The mean number is given by
\begin{eqnarray} \label{RN emission}
{\cal N}_{\mathrm{RN}} = \frac{\sinh(2 \pi \mu_\mathrm{rn}) \sinh(\pi \tilde{\kappa}_\mathrm{rn} - \pi \kappa_\mathrm{rn})}{\cosh(\pi \kappa_\mathrm{rn} + \pi \mu_\mathrm{rn}) \cosh(\pi \tilde{\kappa}_\mathrm{rn} - \pi \mu_\mathrm{rn})}
= \frac{\mathrm{e}^{- 2 \pi (\kappa_\mathrm{rn} - \mu_\mathrm{rn})} - \mathrm{e}^{- 2 \pi (\kappa_\mathrm{rn} + \mu_\mathrm{rn})} }{1 + \mathrm{e}^{- 2 \pi (\kappa_\mathrm{rn} + \mu_\mathrm{rn})}}
\times \frac{1 - \mathrm{e}^{- 2 \pi (\tilde{\kappa}_\mathrm{rn} - \kappa_\mathrm{rn})}}{1 + \mathrm{e}^{- 2 \pi (\tilde{\kappa}_\mathrm{rn} - \mu_\mathrm{rn})}},
\end{eqnarray}
where
\begin{eqnarray}
\kappa_\mathrm{rn} = q Q_0 \frac{r_{\mathrm{ads}}^2}{r_0^2}, \qquad \tilde{\kappa}_\mathrm{rn} = \frac{\omega}{B} r_{\mathrm{ads}}^2, \qquad \mu_\mathrm{rn}^2 = \kappa_\mathrm{rn}^2 - r_{\mathrm{ads}}^2 \bar{m}_\mathrm{rn}^2,
\end{eqnarray}
where $\bar{m}_\mathrm{rn}^2 = m^2 + l(l + 1)/r_0^2 + 1/4 r_\mathrm{ads}^2$. It is interesting to note that~\eqref{RN emission} is formally the inverse of~\eqref{eq_Nout}. The reciprocal relation has been similarly observed in the Schwinger formulae by a constant electric field in $\mathrm{dS}_2$ and $\mathrm{AdS}_2$ space~\cite{Kim:2022nsx}. The physics behind the reciprocal relation requires a further study.

\begin{figure}
\includegraphics[scale=0.12, angle=0]{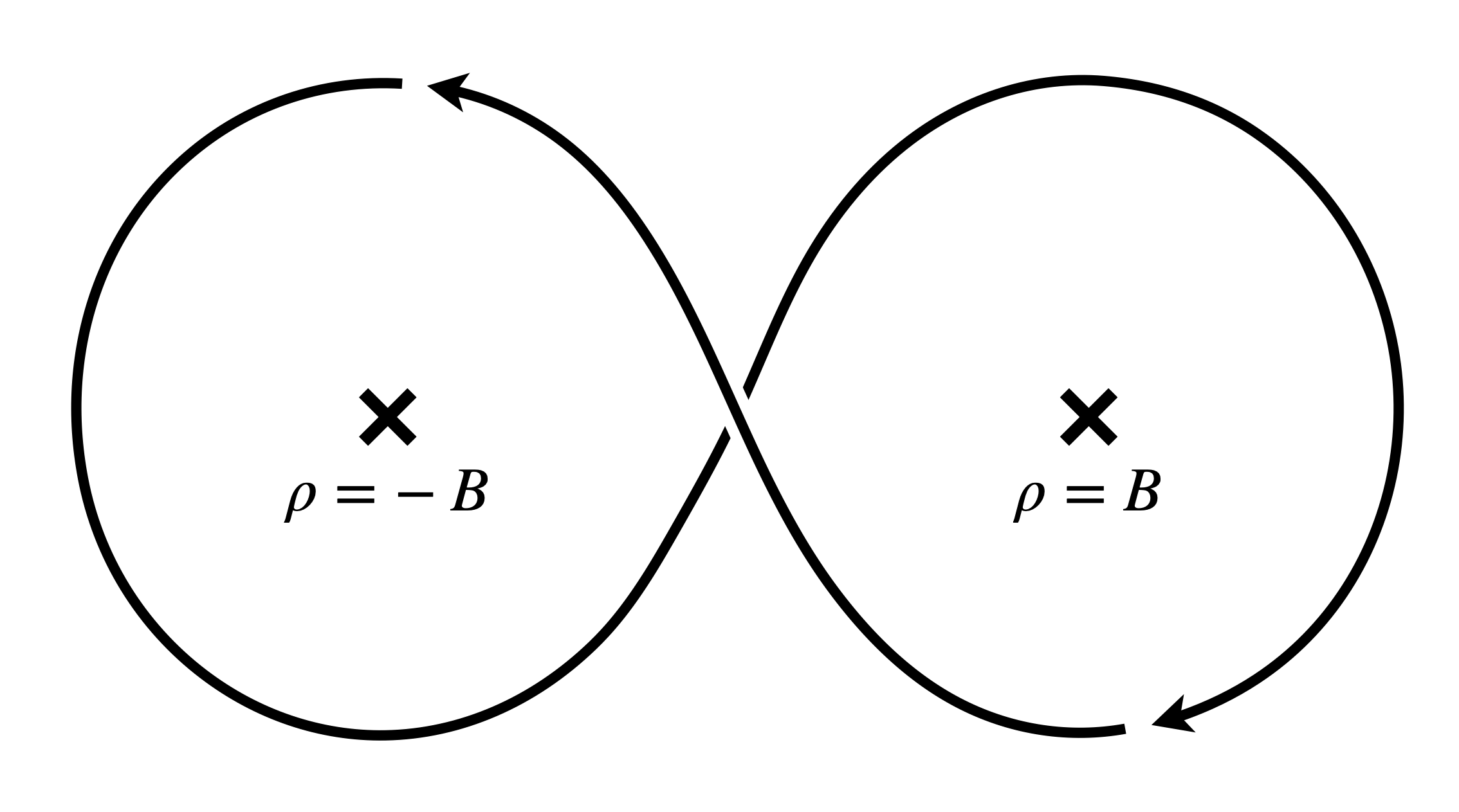}
\caption{The contour of the dominated contribution in phase-integral formulation.}
\label{fig_contour}
\end{figure}

The big difference of the mean number between the Nariai black hole and the near-extremal RN-dS black hole originates from the residues at $\rho = -B$ and $\rho = B$ in the phase-integral formulation~\cite{Kim:2013cka}. The leading Boltzmann factor, for the S-wave ($l = 0$), comes from the contour integral in the complex plane $z$ of $\rho$,
\begin{eqnarray}
{\cal N} \approx \exp\left( i \oint dz \frac{K(z)}{\pm (B^2 - z^2)} \right), \qquad K(z) = \sqrt{(\ell^2 q \mathcal{Q} z/r_h^2 - \omega)^2 \mp \ell^2 m^2 (B^2 - z^2)},
\end{eqnarray}
where the upper (lower) sign for Nariai (RN-dS) black hole, $\mathcal{Q} = Q_n (Q_0), \, \ell = r_{\mathrm{ds}} (r_{\mathrm{ads}}), \, r_h = r_n (r_0)$, and a proper branch-cut outside of the contour is taken to make the square root an analytical function. Then, the contour integral along Figure~\ref{fig_contour} gives ${\cal N} \approx \mathrm{e}^{\pm 2 \pi (K(-B) + K(B))/2B}$ that multiplies to $\mathrm{e}^{\mp 2 \pi \mu}$ from the pole at $z = \infty$, which results in $\mathrm{e}^{\pm 2 \pi (\tilde{\kappa} - \mu)}$. Thus the dS space is the origin of catastrophic emission of charges. Physically the black hole horizon emits charges of the same sign of $\mathcal{Q}$ while the cosmological horizon emits the opposite charges since the electric field points in the opposite direction there.

A few comments are in order.
First, notice that the horizon radius $r_n$ in~(\ref{eq_NariaiLimit}) is larger than $r_0$ in~(\ref{RN-radii}) whereas the dS radius $r_\mathrm{ds}$ in~\eqref{eq_rds} is larger than the AdS radius $r_\mathrm{ads}$ in~\eqref{RN-radii}.
Second, the AdS geometry gives a positive term proportional to the $- R^{(2)}_{\mathrm{ads}}$ to the effective mass for a scalar field whereas the dS geometry subtracts a term proportional to $R^{(2)}_{\mathrm{ds}}$. This is a general feature of pair production in the dS and the AdS space: the effective mass is smaller in the dS space than that in the AdS space. Third, the effective temperature for the leading Boltzmann factor is higher for the dS space with an additional term from the Gibbons-Hawking temperature than the one in the AdS space, in which the curvature term due to the BF bound subtracts the Unruh temperature. These effects are combined to enhance the pair production in the dS space. There is no BF bound for the Nariai black hole, which is the characteristic feature of QED in the dS space.
Finally, the mean number for charge emission in the spacelike region of Nariai black hole has the same structure as the Schwinger formula in planar coordinates of the dS space~\cite{Cai:2014qba}, whose out-vacuum corresponds to the future infinity. The Schwinger formula in the static coordinates of dS space can be obtained from those in Sec.~\ref{sec III} by taking the limit of $r_n = 0, B = 0$ while keeping $Q_n/r_n^2 = E$ and $\tilde{\kappa} = \omega/H$, and then we have $\kappa = q E/H^2$ and $\mu^2 = (q E/H^2)^2 + (m/H)^2$ where $H = 1/r_\mathrm{ds}$.

%%%%%%%%%%%%%%%%%%%%%%%%%%%%%%%%%%%%%%%%%%%%%%%%%%%%%%%%%%%%%%%%%%%%%%
\section{Conclusion}
%%%%%%%%%%%%%%%%%%%%%%%%%%%%%%%%%%%%%%%%%%%%%%%%%%%%%%%%%%%%%%%%%%%%%%

We have studied the emission of charges from charged Nariai black holes. The charged Nariai black hole is the coincident limit of the black hole horizon and the cosmological horizon. We have used the near-extremal charged Nariai black hole whose black hole horizon and cosmological horizon are separated by a distance smaller than the black hole horizon. The electric field from the charge of the black hole points radially, and thus the two horizons play the role of a spherical conductor. The Hawking temperature for the black hole horizon and the Gibbons-Hawking temperature for the cosmological horizon decrease proportional to the separation, and thereby the radiations from both horizons are exponentially suppressed. However, one may expect from the Schwinger effect that charges of the same sign as the black hole will be spontaneously emitted from the black hole horizon, but charges of the opposite sign will be produced from the cosmological horizon and then fall to the black hole, which will speed up the discharge process.

To quantify the Schwinger formula for spontaneously produced pairs from both horizons, we have used the enhanced symmetry of the near-extremal charged Nariai black hole, which has the near-horizon geometry of $\mathrm{dS}_2 \times \mathrm{S}^2$, one of the geometry for the Einstein-Maxwell theory. The charged scalar field has solutions in the timelike region between two horizons and another solutions in the spacelike region beyond the cosmological horizon, from which we properly find the in- and out-going modes in the asymptotic regions in the timelike region and the spacelike region, respectively, and then calculate the Bogoliubov coefficients between the in- and the out-vacua.

The mean number of spontaneously produced charges, i.e, the emission, exhibits exponentially enhanced emission in between two horizons. This catastrophic emission of exponentially exploding number of charges with energy greater than their chemical potentials is a consequence of the existence of two close horizons in the dS space, which strongly contrasts to two horizons of a near-extremal RN-dS black hole with the near-horizon geometry of $\mathrm{AdS}_2 \times \mathrm{S}^2$. We argue that the dS space with two close horizons results in the catastrophic emission whereas the AdS space with two close horizons results in an exponentially bounded emission. However, a near-extremal charged Nariai black hole can end up to two different spacetimes depending on the charge-mass ratio of emitted particles. By emitting more heavy particles, the black hole loses more mass than charge and becomes a nonextremal RN-dS black hole. Conversely, if the black hole emits light particles then, by losing more charge, it ends up to another black hole with one horizon and a naked singularity. The ``sufficient'' condition to avoid the formation of naked singularity is $m > \sqrt2 q$, which can be obtained from the minimum slope of $dQ/dM$ (the maximum slope of $dM/dQ$) of the curve of $r_+ = r_c$ at the ultracold point.

It will be interesting to further study the exact evolution of Nariai black holes due to the pair production. It should be noted that the Schwinger emission of charges from charged Nariai black holes is a result of the vacuum persistence amplitude, twice the imaginary part of one-loop effective action for a charged scalar field, which is a consequence of the vacuum instability due to pair production~\cite{Kim:2008yt,Kim:2009pg}.
In fact, the mean number ${\cal N}$ of produced scalar charges for a given quantum number is related to the vacuum persistence amplitude as $\ln (1 + {\cal N})$. Thus, the relevant framework for the evolution of the charged Nariai black holes due to the emission of charges is to include the energy-momentum tensors from the Maxwell action and the one-loop effective action and also from the induced four-current from produced charges in the Maxwell field~\cite{Ruffini:2009hg}. This issue goes beyond the scope of the present work and will be addressed in a future work.

Another interesting direction is to consider the model with a dynamical ``cosmological constant'', for instance, a scalar field for inflation. The Einstein-Maxwell theory including one-loop effects keeps the cosmological constant or the dS radius as a fixed parameter. However, the Einstein-Maxwell theory coupled to a complex scalar field as dynamical cosmological constant is a legitimate model to describe the decay of the cosmological constant through the emission of charged pairs. This issue is also beyond the scope of present work and will be investigated in the future.

\acknowledgments
The authors would like to thank Hyun Kyu Lee for helpful discussions, and Miguel Montero, Thomas Van Riet and Gerben Venken for useful comments on the back-reaction of pair production and are grateful to the anonymous referee for helpful comments on the evolution of extremal black holes.
C.M.C. would like to thank the warm hospitality at Kunsan National University and Center for Quantum Spacetime (CQUeST), Sogang University, where this work was initiated.
S.P.K. would like to appreciate the warm hospitality at CQUeST and ELI Beamlines, Czech Republic, where part of this work was done and Center for High Energy and High Field Physics (CHiP), National Central University, where this work was revised.
The work of C.M.C. was supported by the National Science and Technology Council of the R.O.C. (Taiwan) under the grants NSTC 111-2112-M-008-012, 112-2112-M-008-020.
The work of S.P.K. was supported in part by National Research Foundation of Korea (NRF) funded by the Ministry of Education (2019R1I1A3A01063183).

\begin{appendix}

%%%%%%%%%%%%%%%%%%%%%%%%%%%%%%%%%%%%%%%%%%%%%%%%%%%%%%%%%%%%%%%%%%%%%%
\section{Boundary Conditions for Tunneling and Scattering Processes} \label{app_A}
%%%%%%%%%%%%%%%%%%%%%%%%%%%%%%%%%%%%%%%%%%%%%%%%%%%%%%%%%%%%%%%%%%%%%%
The boundary conditions on a quantum field for pair production differ in the timelike inner region and the spacelike outer region. In the inner region the quantum field describes a tunneling process and the mean number for pair production is determined by the flux ratio while in the outer region the quantum field scatters over a potential barrier and the mean number is given by the ratio of the out-going negative frequency to the in-going positive frequency, i.e, the energy flow ratio.

\begin{figure}
%\subfloat[]
\includegraphics[scale=0.15, angle=0]{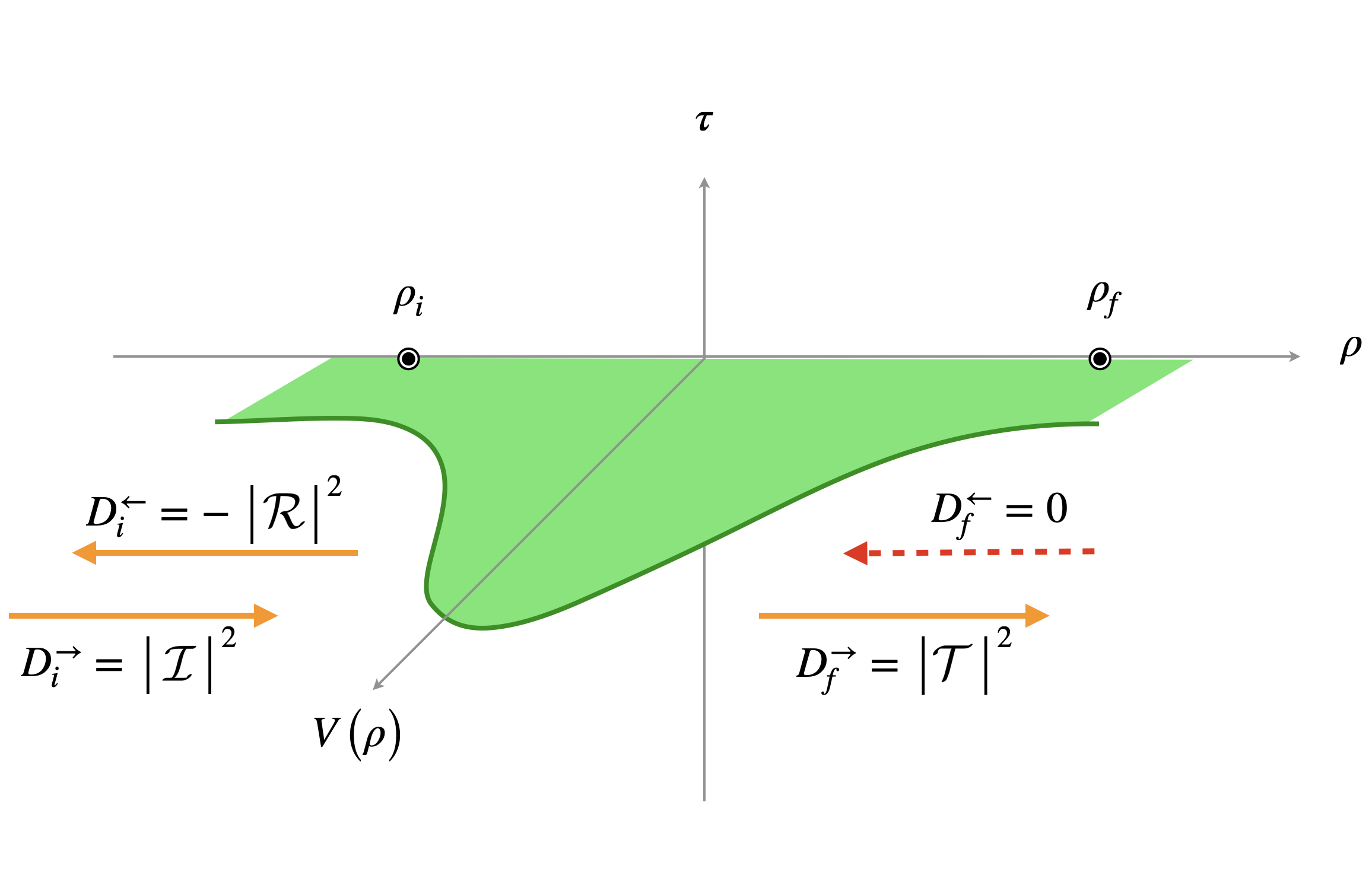}
\hspace{.5cm}
\includegraphics[scale=0.17, angle=0]{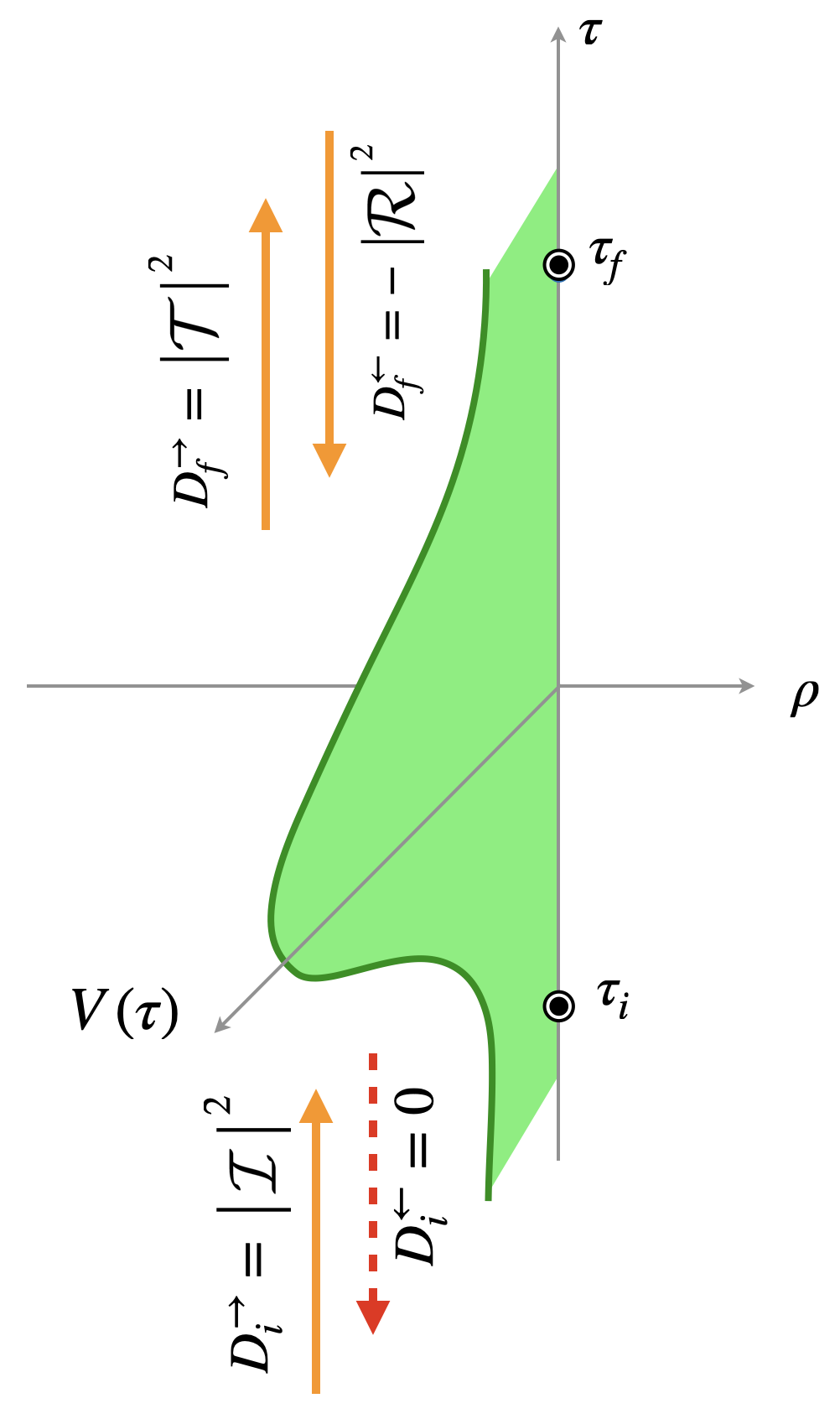}
\caption{Boundary conditions for tunneling [left] and scattering [right] processes.}
\label{fig_BCs}
\end{figure}

For the pair production in the inner region the spacetime is static, and the mode equation of KG equation reduces to a second order ordinary differential equation with respect to the radial coordinate $\rho$. It describes a tunneling process as shown in the left panel of Fig.~\ref{fig_BCs}. As discussed in~\cite{Chen:2012zn}, we impose the zero in-going mode (left moving mode) flux $D_f^\leftarrow = 0$ at $\rho_f$, an asymptotically far right region. Then the physical interpretation for the other three fluxes are as follows. The out-going mode (right moving mode) flux  $D_f^\rightarrow$ at $\rho_f$ corresponds to produced particles (also named as transmitted flux $|\mathcal{T}|^2$), the out-going mode flux $D_i^\rightarrow $ at $\rho_i$, an asymptotically far left region, corresponds to virtual particles by quantum fluctuations from vacuum (also named as incident flux $|\mathcal{I}|^2$), and the in-going mode flux $D_i^\leftarrow $ at $\rho_i$ denotes the re-annihilation part of virtual particles (also named as reflected flux $|\mathcal{R}|^2$). These fluxes are conserved (the out-going mode flux is positive and the in-mode is negative)
\begin{equation}
| D_i^\rightarrow | = | D_i^\leftarrow | + | D_f^\rightarrow | \quad \Rightarrow \quad D_i^\rightarrow + D_i^\leftarrow = D_f^\rightarrow.
\end{equation}
The mean number is defined by the ratio of the out-going flux at $\rho_f$ to the in-going flux at $\rho_i$, namely,
\begin{equation}
\mathcal{N}_\mathrm{tunneling} = \frac{| D_f^\rightarrow |}{| D_i^\leftarrow |} = - \frac{D_f^\rightarrow}{D_i^\leftarrow} = \frac{|\mathcal{T}|^2}{|\mathcal{R}|^2}.
\end{equation}

However, the physical process of the pair production in the outer region is different. In this region, the Killing vector $\partial / \partial t$ becomes spacelike and the spacetime becomes time-dependent, in fact, an expanding geometry. There the KG equation reduces to a second order ordinary differential equation with respect to a timelike coordinate $\tau$. It indeed describes a scattering process as shown in the right panel of Fig.~\ref{fig_BCs}. For this case, we impose the zero in-going mode (backward mode in time or negative frequency mode) $D_i^\leftarrow = 0$ at $\tau_i$, the past infinity, and then the physical interpretation for the other three energy densities are: the out-going mode (forward mode in time or positive frequency mode) $D_i^\rightarrow = |\mathcal{I}|^2$ at $\tau_i$ corresponds to incident particles, the out-going mode $D_f^\rightarrow = |\mathcal{T}|^2$ at $\tau_f$, the future infinity, are transmitted particles, and the in-going mode $D_f^\leftarrow = - |\mathcal{R}|^2$ at $\tau_f$ denotes the produced particles. The conservation of energy becomes
\begin{equation}
| D_i^\rightarrow | = | D_f^\rightarrow | - | D_f^\leftarrow | \quad \Rightarrow \quad D_i^\rightarrow = D_f^\rightarrow + D_f^\leftarrow.
\end{equation}
The mean number is defined to describe the ratio of the produced particles to the incident particles, namely,
\begin{equation}
\mathcal{N}_\mathrm{scattering} = \frac{| D_f^\leftarrow |}{| D_i^\rightarrow |} = - \frac{D_f^\leftarrow}{D_i^\rightarrow}  = \frac{|\mathcal{R}|^2}{|\mathcal{I}|^2}.
\end{equation}

%%%%%%%%%%%%%%%%%%%%%%%%%%%%%%%%%%%%%%%%%%%%%%%%%%%%%%%%%%%%%%%%%%%%%%
\section{General Formula for Mean Number} \label{app_B}
%%%%%%%%%%%%%%%%%%%%%%%%%%%%%%%%%%%%%%%%%%%%%%%%%%%%%%%%%%%%%%%%%%%%%%
For self-containment, we recapitulate the monodromy method to calculate the mean number for the Schwinger effect~\cite{Chen:2022hpe}. Let us consider the Riemann differential equation
\begin{eqnarray} \label{eq_Riemann}
\frac{d^2 \Phi(z)}{dz^2} &+& \left( \frac{1 - \alpha_1 - \beta_1}{z - z_1} + \frac{1 - \alpha_2 - \beta_2}{z - z_2} \right) \frac{d \Phi(z)}{dz}
\nonumber\\
&+& \frac{1}{(z - z_1)(z - z_2)} \left( \frac{\alpha_1 \beta_1 (z_1 - z_2)}{z - z_1} + \frac{\alpha_2 \beta_2 (z_2 - z_1)}{z - z_2} + \alpha_\infty \beta_\infty \right) \Phi(z) = 0,
\end{eqnarray}
where the characteristic exponents satisfy the condition $\alpha_1 + \beta_1 + \alpha_2 + \beta_2 + \alpha_\infty + \beta_\infty = 1$. The hypergeometric equations in Sec.~\ref{sec III} are special cases of~\eqref{eq_Riemann}. The solution can be expressed as a P-function
\begin{equation} \label{eq_Pfun}
P \begin{pmatrix} z_1 & z_2 & z_\infty & \\ \alpha_1 & \alpha_2 & \alpha_\infty & ; z \\ \beta_1 & \beta_2 & \beta_\infty & \end{pmatrix},
\end{equation}
where to implement the pair production from $z = z_2$ to $z = z_\infty$, the imaginary parts of $\alpha_2$ and $\alpha_\infty$ (similarly, the sign of $\beta_2$ and $\beta_\infty$) should have the same sign. The associated monodromy matrices are
\begin{eqnarray}
&& \mathbf{M}_1 = \begin{bmatrix} \mathrm{e}^{2 \pi i \alpha_1} & 1 \\ 0 & \mathrm{e}^{2 \pi i \beta_1} \end{bmatrix}, \qquad \mathbf{M}_2 = \begin{bmatrix} \mathrm{e}^{2 \pi i \alpha_2} & 0 \\ b & \mathrm{e}^{2 \pi i \beta_2} \end{bmatrix},
\nonumber\\
&& \mathbf{M}_\infty = \begin{bmatrix} \mathrm{e}^{- 2 \pi i (\alpha_1 + \alpha_2)} & - \mathrm{e}^{- 2 \pi i (\alpha_1 + \alpha_2 + \beta_1)} \\ - b \, \mathrm{e}^{- 2 \pi i (\alpha_1 + \alpha_2 + \beta_2)} & \mathrm{e}^{- 2 \pi i (\beta_1 + \beta_2)} + b \, \mathrm{e}^{- 2 \pi i (\alpha_1 + \alpha_2 + \beta_1 + \beta_2)} \end{bmatrix},
\end{eqnarray}
where
\begin{equation}
b = \mathrm{e}^{-2 \pi i \alpha_\infty} + \mathrm{e}^{-2 \pi i \beta_\infty} - \mathrm{e}^{2 \pi i (\alpha_1 + \alpha_2)} - \mathrm{e}^{2 \pi i (\beta_1 + \beta_2)}.
\end{equation}
The eigenvalues of $\mathbf{M}_2, \mathbf{M}_\infty$ are $(\mathrm{e}^{2 \pi i \alpha_2}, \mathrm{e}^{2 \pi i \beta_2}), (\mathrm{e}^{2 \pi i \beta_\infty}, \mathrm{e}^{2 \pi i \alpha_\infty})$, respectively, and their eigenvectors composing matrices are
\begin{equation}
\mathbf{E}_2 = \begin{bmatrix} \mathrm{e}^{2 \pi i \alpha_2} - \mathrm{e}^{2 \pi i \beta_2} & 0 \\ b & 1 \end{bmatrix}, \qquad \mathbf{E}_\infty = \begin{bmatrix} \mathrm{e}^{2 \pi i \beta_2} & \mathrm{e}^{2 \pi i \beta_2} \\ \mathrm{e}^{2 \pi i (\beta_1 + \beta_2)} - \mathrm{e}^{- 2 \pi i \alpha_\infty} & \mathrm{e}^{2 \pi i (\beta_1 + \beta_2)} - \mathrm{e}^{- 2 \pi i \beta_\infty} \end{bmatrix}.
\end{equation}
Consequently, the connection matrix relating $z = z_2$ to $z = z_\infty$ is
\begin{eqnarray} \label{eq_connec mat}
\mathbf{P}_2^\infty &=& \begin{bmatrix} d_1 & 0 \\ 0 & d_2 \end{bmatrix} (\mathbf{E}_2)^{-1} \mathbf{E}_\infty \begin{bmatrix} d_3 & 0 \\ 0 & d_4 \end{bmatrix} = \frac1{\mathrm{e}^{2 \pi i \alpha_2} - \mathrm{e}^{2 \pi i \beta_2}} \begin{bmatrix} d_1 d_3 \, \mathrm{e}^{2 \pi i \beta_2} & d_1 d_4 \, \mathrm{e}^{2 \pi i \beta_2} \\ - d_2 d_3 \, \Xi_1 & - d_2 d_4 \, \Xi_2 \end{bmatrix},
\\
&& \Xi_1 = \mathrm{e}^{2 \pi i (\alpha_2 - \alpha_\infty)} + \mathrm{e}^{2 \pi i (\beta_2 - \beta_\infty)}  - \mathrm{e}^{2 \pi i (\alpha_1 + \alpha_2 + \beta_2)} - \mathrm{e}^{2 \pi i (\beta_1 + \beta_2 + \alpha_2)},
\nonumber\\
&& \Xi_2 = \mathrm{e}^{2 \pi i (\alpha_2 - \beta_\infty)} + \mathrm{e}^{2 \pi i (\beta_2 - \alpha_\infty)} - \mathrm{e}^{2 \pi i (\alpha_1 + \alpha_2 + \beta_2)} - \mathrm{e}^{2 \pi i (\beta_1 + \beta_2 + \alpha_2)}.
\end{eqnarray}
The unit determinant of the connection matrix $\mathbf{P}_2^\infty$
\begin{eqnarray}
d_1 d_2 d_3 d_4
%&=& - \frac{\left( \mathrm{e}^{2 \pi i \alpha_2} - \mathrm{e}^{2 \pi i \beta_2} \right)^2}{ \mathrm{e}^{2 \pi i \beta_2} \left[ \mathrm{e}^{2 \pi i (\alpha_2 - \beta_\infty)} - \mathrm{e}^{2 \pi i (\alpha_2 - \alpha_\infty)} + \mathrm{e}^{2 \pi i (\beta_2 - \alpha_\infty)} - \mathrm{e}^{2 \pi i (\beta_2 - \beta_\infty)} \right]} \nonumber\\
&=& \frac{\mathrm{e}^{2 \pi i \alpha_2} - \mathrm{e}^{2 \pi i \beta_2}}{\mathrm{e}^{2 \pi i \beta_2} \left( \mathrm{e}^{- 2 \pi i \alpha_\infty} - \mathrm{e}^{- 2 \pi i \beta_\infty} \right)},
\end{eqnarray}
ensures the conservation of energy/flux.

The physical correspondence of the connection matrix~\eqref{eq_connec mat} is different for tunneling and scattering processes. For the tunneling process the components of the connection matrix are related to the fluxes for pair production as~\cite{Chen:2022hpe}
\begin{equation}
\mathbf{P}_2^\infty = \begin{bmatrix} \mathcal{I}/\mathcal{T} & \mathcal{R}/\mathcal{T} \\ \mathcal{R}^*/\mathcal{T}^* & \mathcal{I}^*/\mathcal{T}^* \end{bmatrix}.
\end{equation}
Therefore the mean number of pair production via the tunneling process is
\begin{eqnarray} \label{eq_Nt}
N_\textrm{tunneling} = \frac{|\mathcal{T}|^2}{|\mathcal{R}|^2} &=& \frac{\mathrm{e}^{2 \pi i (\alpha_2 - \beta_\infty)} - \mathrm{e}^{2 \pi i (\alpha_2 - \alpha_\infty)} + \mathrm{e}^{2 \pi i (\beta_2 - \alpha_\infty)} - \mathrm{e}^{2 \pi i (\beta_2 - \beta_\infty)}}{\mathrm{e}^{2 \pi i (\alpha_2 - \alpha_\infty)} + \mathrm{e}^{2 \pi i (\beta_2 - \beta_\infty)}  - \mathrm{e}^{2 \pi i (\alpha_1 + \alpha_2 + \beta_2)} - \mathrm{e}^{2 \pi i (\beta_1 + \beta_2 + \alpha_2)}} \nonumber\\
&=& \frac{\sin \pi (\alpha_2 - \beta_2) \sin \pi(\beta_\infty - \alpha_\infty)}{\sin \pi(\alpha_1 + \beta_2 + \alpha_\infty) \sin \pi(\alpha_1 + \alpha_2 + \beta_\infty)}.
\end{eqnarray}
Here the constraint $\alpha_1 + \beta_1 + \alpha_2 + \beta_2 + \alpha_\infty + \beta_\infty = 1$ has been used.

However, for the scattering process, following the argument in~\cite{Chen:2022hpe}, the correspondence between components of connection matrix and energy flows becomes
\begin{equation}
\mathbf{P}_2^\infty = \begin{bmatrix} \mathcal{T}^*/\mathcal{I}^* & - \mathcal{R}^*/\mathcal{I}^* \\ -\mathcal{R}/\mathcal{I} & \mathcal{T}/\mathcal{I} \end{bmatrix},
\end{equation}
and the associated the mean number is
\begin{equation} \label{eq_Ns}
N_\textrm{scattering} = \frac{|\mathcal{R}|^2}{|\mathcal{I}|^2} = \frac{\sin \pi(\alpha_1 + \beta_2 + \alpha_\infty) \sin \pi(\alpha_1 + \alpha_2 + \beta_\infty)}{\sin \pi (\alpha_2 - \beta_2) \sin \pi(\beta_\infty - \alpha_\infty)} = \frac1{N_\textrm{tunneling}}.
\end{equation}

The mean number formulae~\eqref{eq_Nt} and~\eqref{eq_Ns} consider the pair production from $z_2$ to $z_\infty$. With a suitable permutation of $(\alpha_i, \beta_i)$ one can compute the mean number of pair production in different regions. It is worth to emphasize that these formulae are very helpful in computing the mean number directly from the KG equation.

\end{appendix}

%%%%%%%%%%%%%%%%%%%%%%%%%%%%%%%%%%%%%%%%%%%%%%%%%%%%%%%%%%%%%%%%%%%%%%


\begin{thebibliography}{99}
%%%%%%%%%%%%%%%%%%%%%%%%%%%%%%%%%%%%%%%%%%%%%%%%%%%%%%%%%%%%%%%%%%%%%%

%\cite{Gibbons:1977mu}
\bibitem{Gibbons:1977mu}
G.~W.~Gibbons and S.~W.~Hawking,
``Cosmological Event Horizons, Thermodynamics, and Particle Creation,''
Phys. Rev. D \textbf{15} (1977), 2738-2751
doi:10.1103/PhysRevD.15.2738

%\cite{Gregory:2021ozs}
\bibitem{Gregory:2021ozs}
R.~Gregory, I.~G.~Moss, N.~Oshita and S.~Patrick,
``Black hole evaporation in de Sitter space,''
Class. Quant. Grav. \textbf{38} (2021) no.18, 185005
doi:10.1088/1361-6382/ac1a68
[arXiv:2103.09862 [gr-qc]].

%\cite{Romans:1991nq}
\bibitem{Romans:1991nq}
L.~J.~Romans,
``Supersymmetric, cold and lukewarm black holes in cosmological Einstein-Maxwell theory,''
Nucl. Phys. B \textbf{383} (1992), 395-415
doi:10.1016/0550-3213(92)90684-4
[arXiv:hep-th/9203018 [hep-th]].

%\cite{Bardeen:1999px}
\bibitem{Bardeen:1999px}
J.~M.~Bardeen and G.~T.~Horowitz,
``The Extreme Kerr throat geometry: A Vacuum analog of $\mathrm{AdS}_2 \times \mathrm{S}^2$,''
Phys. Rev. D \textbf{60} (1999), 104030
doi:10.1103/PhysRevD.60.104030
[arXiv:hep-th/9905099 [hep-th]].


%\cite{Chen:2020mqs}
\bibitem{Chen:2020mqs}
C.-M.~Chen and S.~P.~Kim,
``Schwinger Effect from Near-extremal Black Holes in (A)dS Space,''
Phys. Rev. D \textbf{101} (2020) no.8, 085014
doi:10.1103/PhysRevD.101.085014
[arXiv:2002.00394 [hep-th]].

%\cite{Chen:2021jwy}
\bibitem{Chen:2021jwy}
C.-M.~Chen and S.~P.~Kim,
``Dyon production from near-extremal Kerr-Newman-(anti-)de Sitter black holes,''
Eur. Phys. J. C \textbf{83} (2023) no.3, 219
doi:10.1140/epjc/s10052-023-11380-7
[arXiv:2111.14124 [hep-th]].


%\cite{Castro:2022cuo}
\bibitem{Castro:2022cuo}
A.~Castro, F.~Mariani and C.~Toldo,
``Near-extremal limits of de Sitter black holes,''
JHEP \textbf{07} (2023), 131
doi:10.1007/JHEP07(2023)131
[arXiv:2212.14356 [hep-th]].

%\cite{Garriga:1994bm}
\bibitem{Garriga:1994bm}
J.~Garriga,
``Pair production by an electric field in (1+1)-dimensional de Sitter space,''
Phys. Rev. D \textbf{49} (1994), 6343-6346
doi:10.1103/PhysRevD.49.6343

%\cite{Kim:2008xv}
\bibitem{Kim:2008xv}
S.~P.~Kim and D.~N.~Page,
``Schwinger Pair Production in dS$_2$ and AdS$_2$,''
Phys. Rev. D \textbf{78} (2008), 103517
doi:10.1103/PhysRevD.78.103517
[arXiv:0803.2555 [hep-th]].

%\cite{Pioline:2005pf}
\bibitem{Pioline:2005pf}
B.~Pioline and J.~Troost,
``Schwinger pair production in AdS$_2$,''
JHEP \textbf{03} (2005), 043
doi:10.1088/1126-6708/2005/03/043
[arXiv:hep-th/0501169 [hep-th]].

%\cite{Ortaggio:2002bp}
\bibitem{Ortaggio:2002bp}
M.~Ortaggio and J.~Podolsky,
``Impulsive waves in electrovac direct product space-times with Lambda,''
Class. Quant. Grav. \textbf{19} (2002), 5221-5227
doi:10.1088/0264-9381/19/20/313
[arXiv:gr-qc/0209068 [gr-qc]].

%\cite{Chen:2022hpe}
\bibitem{Chen:2022hpe}
C.-M.~Chen, T.~Ishige, S.~P.~Kim, A.~Takayasu and C.-Y.~Wei,
``Monodromy approach to pair production of charged black holes and electric fields,''
Chin. J. Phys. \textbf{86}, 255-268 (2023)
doi:10.1016/j.cjph.2023.10.007
[arXiv:2210.14792 [hep-th]].

%\cite{Castro:2013kea}
\bibitem{Castro:2013kea}
A.~Castro, J.~M.~Lapan, A.~Maloney and M.~J.~Rodriguez,
``Black Hole Monodromy and Conformal Field Theory,''
Phys. Rev. D \textbf{88} (2013), 044003
doi:10.1103/PhysRevD.88.044003
[arXiv:1303.0759 [hep-th]].

%\cite{Castro:2013lba}
\bibitem{Castro:2013lba}
A.~Castro, J.~M.~Lapan, A.~Maloney and M.~J.~Rodriguez,
``Black Hole Scattering from Monodromy,''
Class. Quant. Grav. \textbf{30} (2013), 165005
doi:10.1088/0264-9381/30/16/165005
[arXiv:1304.3781 [hep-th]].

%\cite{Montero:2019ekk}
\bibitem{Montero:2019ekk}
M.~Montero, T.~Van Riet and G.~Venken,
``Festina Lente: EFT Constraints from Charged Black Hole Evaporation in de Sitter,''
JHEP \textbf{01}(2020), 039
doi:10.1007/JHEP01(2020)039
[arXiv:1910.01648 [hep-th]].

%\cite{Cai:2014qba}
\bibitem{Cai:2014qba}
R.-G.~Cai and S.~P.~Kim,
``One-Loop Effective Action and Schwinger Effect in (Anti-) de Sitter Space,''
JHEP \textbf{09} (2014), 072
doi:10.1007/JHEP09(2014)072
[arXiv:1407.4569 [hep-th]].

%\cite{Cardoso:2004uz}
\bibitem{Cardoso:2004uz}
V.~Cardoso, O.~J.~C.~Dias and J.~P.~S.~Lemos,
``Nariai, Bertotti-Robinson and anti-Nariai solutions in higher dimensions,''
Phys. Rev. D \textbf{70} (2004), 024002
doi:10.1103/PhysRevD.70.024002
[arXiv:hep-th/0401192 [hep-th]].

%\cite{Costa:2018zvw}
\bibitem{Costa:2018zvw}
J.~L.~Costa, J.~Nat\'ario and P.~Oliveira,
``Cosmic no-hair in spherically symmetric black hole spacetimes,''
Annales Henri Poincare \textbf{20} (2019) no.9, 3059-3090
doi:10.1007/s00023-019-00825-z
[arXiv:1801.06549 [gr-qc]].

%\cite{Kim:2008yt}
\bibitem{Kim:2008yt}
S.~P.~Kim, H.~K.~Lee and Y.~Yoon,
``Effective Action of QED in Electric Field Backgrounds,''
Phys. Rev. D \textbf{78}, 105013 (2008)
doi:10.1103/PhysRevD.78.105013
[arXiv:0807.2696 [hep-th]].

%\cite{Chen:2012zn}
\bibitem{Chen:2012zn}
C.-M.~Chen, S.~P.~Kim, I.-C.~Lin, J.-R.~Sun and M.-F.~Wu,
``Spontaneous Pair Production in Reissner-Nordstrom Black Holes,''
Phys. Rev. D \textbf{85} (2012), 124041
doi:10.1103/PhysRevD.85.124041
[arXiv:1202.3224 [hep-th]].

%\cite{Kim:2009pg}
\bibitem{Kim:2009pg}
S.~P.~Kim, H.~K.~Lee and Y.~Yoon,
``Effective Action of QED in Electric Field Backgrounds II. Spatially Localized Fields,''
Phys. Rev. D \textbf{82}, 025015 (2010)
doi:10.1103/PhysRevD.82.025015
[arXiv:0910.3363 [hep-th]].

%\cite{Frob:2014zka}
\bibitem{Frob:2014zka}
M.~B.~Fr\"ob, J.~Garriga, S.~Kanno, M.~Sasaki, J.~Soda, T.~Tanaka and A.~Vilenkin,
``Schwinger effect in de Sitter space,''
JCAP \textbf{04}, 009 (2014)
doi:10.1088/1475-7516/2014/04/009
[arXiv:1401.4137 [hep-th]].

%\cite{Kobayashi:2014zza}
\bibitem{Kobayashi:2014zza}
T.~Kobayashi and N.~Afshordi,
``Schwinger Effect in 4D de Sitter Space and Constraints on Magnetogenesis in the Early Universe,''
JHEP \textbf{10}, 166 (2014)
doi:10.1007/JHEP10(2014)166
[arXiv:1408.4141 [hep-th]].

%\cite{Bavarsad:2016cxh}
\bibitem{Bavarsad:2016cxh}
E.~Bavarsad, C.~Stahl and S.~S.~Xue,
``Scalar current of created pairs by Schwinger mechanism in de Sitter spacetime,''
Phys. Rev. D \textbf{94}, no.10, 104011 (2016)
doi:10.1103/PhysRevD.94.104011
[arXiv:1602.06556 [hep-th]].

%\cite{Stahl:2016geq}
\bibitem{Stahl:2016geq}
C.~Stahl and S.~S.~Xue,
``Schwinger effect and backreaction in de Sitter spacetime,''
Phys. Lett. B \textbf{760}, 288-292 (2016)
doi:10.1016/j.physletb.2016.07.011
[arXiv:1603.07166 [hep-th]].

%\cite{Banyeres:2018aax}
\bibitem{Banyeres:2018aax}
M.~Banyeres, G.~Dom\`enech and J.~Garriga,
``Vacuum birefringence and the Schwinger effect in (3+1) de Sitter,''
JCAP \textbf{10}, 023 (2018)
doi:10.1088/1475-7516/2018/10/023
[arXiv:1809.08977 [hep-th]].

%\cite{Meimanat:2023hjq}
\bibitem{Meimanat:2023hjq}
O.~G.~Meimanat and E.~Bavarsad,
``Induced energy-momentum tensor in de~Sitter scalar QED and its implication for induced currents,''
Phys. Rev. D \textbf{107}, no.12, 125001 (2023)
doi:10.1103/PhysRevD.107.125001
[arXiv:2301.04227 [hep-th]].

%\cite{Moss:2023kah}
\bibitem{Moss:2023kah}
I.~G.~Moss and P.~Z.~Stasiak,
``Mining the quantum vacuum: quantum tunnelling and particle creation,''
J. Phys. Conf. Ser. \textbf{2531}, no.1, 012014 (2023)
doi:10.1088/1742-6596/2531/1/012014
[arXiv:2303.01119 [hep-th]].

%\cite{Kim:2022nsx}
\bibitem{Kim:2022nsx}
S.~P.~Kim, W.-Y.~P.~Hwang and T.-C.~Wang,
``Schwinger mechanism in global $dS_2$ and $AdS_2$ space,''
Chin. J. Phys. \textbf{77} (2022), 2073-2077
doi:10.1016/j.cjph.2022.01.004

%\cite{Kim:2013cka}
\bibitem{Kim:2013cka}
S.~P.~Kim,
``Geometric Origin of Stokes Phenomenon for de Sitter Radiation,''
Phys. Rev. D \textbf{88} (2013) no.4, 044027
doi:10.1103/PhysRevD.88.044027
[arXiv:1307.0590 [hep-th]].

%\cite{Ruffini:2009hg}
\bibitem{Ruffini:2009hg}
R.~Ruffini, G.~Vereshchagin and S.~S.~Xue,
``Electron-positron pairs in physics and astrophysics: from heavy nuclei to black holes,''
Phys. Rept. \textbf{487} (2010) 1-140
doi:10.1016/j.physrep.2009.10.004
[arXiv:0910.0974 [astro-ph.HE]].

\end{thebibliography}
\end{document}